\documentclass[final,twoside]{IEEEtran} 

\usepackage{float}
\usepackage{amssymb,amsmath,amsthm,bm}

\makeatletter
\newcommand{\tpmod}[1]{{\@displayfalse\pmod{#1}}}
\makeatother

\usepackage{graphicx,color}

\graphicspath{{figures-pdf/}}
\usepackage{cite}
\usepackage{url}
\usepackage{diagbox}
\usepackage[aboveskip=1pt]{subcaption}

\usepackage{algorithm}
\usepackage{algpseudocode}
\usepackage{multirow,bigstrut}
\usepackage{tabularx}
\usepackage{arydshln}
\usepackage{empheq}
\usepackage{datetime}

\newlength\OneImW
\newlength\TwoImW
\newlength\twofigwidth
\newlength\sfigwidth
\newlength\vfigskip
\newlength\imagewidth
\newlength\figsep
\newcommand{\vct}[1]{\bm{#1}}
\newcommand{\mtx}[1]{\bm{#1}}

\newcommand{\Fee}{\mtx{\Phi}}

\newcommand{\norm}[1]{\left\Vert {#1} \right\Vert}
\newcommand{\pnorm}[2]{\norm{#2}_{#1}}

\usepackage[bookmarks=false]{hyperref}
\hypersetup{
 linktocpage=true, pdfborderstyle={/S/S/W 1}, hyperindex=true, bookmarks=true, bookmarksopen=true, bookmarksnumbered=true,
}

\makeatother

\begin{document}

\title{Multi-Channel Deep Networks for Block-Based Image Compressive Sensing}

\author{Siwang~Zhou,~\IEEEmembership{Member,~IEEE,}
	Yan~He,
	Yonghe~Liu,
	Chengqing~Li,~\IEEEmembership{Senior Member,~IEEE,}
    and~Jianming~Zhang


\thanks{This work was supported in part by the National Science Foundation of China under Grant 61772447, 61972056, in part by the Hunan Provincial National Science Foundation under Grant 2019JJ40047. \emph{(Corresponding author: Siwang Zhou.)}}

\thanks{S. Zhou, Y. He, and C. Li are with the College of Computer Science and Electrical Engineering, Hunan University, Changsha 410082, China (email: \{swzhou, heyan, chengqingli\}@hnu.edu.cn).}

\thanks{Y. Liu is with the Department of Computer Science and Engineering, the University of Texas at Arlington, TX 76019, USA (email: yonghe@cse.uta.edu).}

\thanks{J. Zhang is with the School of Computer and Communication Engineering, Changsha University of Science and Technology, Changsha 410114, China (email: jmzhang@csust.edu.cn).}}

\markboth{IEEE Transactions}{Zhou \MakeLowercase{\textit{et al.}}}
	
\IEEEpubid{\begin{minipage}{\textwidth}\ \\[12pt] \centering
			1520-9210 \copyright 2020 IEEE. Personal use is permitted, but republication/redistribution requires IEEE permission.\\
			See https://www.ieee.org/publications/rights/index.html for more information.
	\end{minipage}
}

\maketitle

\begin{abstract}
Incorporating deep neural networks in image compressive sensing (CS) receives intensive attentions in multimedia technology and applications recently. As deep network approaches learn the inverse mapping directly from the CS measurements, the reconstruction speed is significantly faster than the conventional CS algorithms. However, for existing network-based approaches, a CS sampling procedure has to map a separate network model. This may potentially degrade the performance of image CS with block-wise sampling because of blocking artifacts, especially when multiple sampling rates are assigned to different blocks within an image. In this paper, we develop a multi-channel deep network for block-based image CS by exploiting inter-block correlation with performance significantly exceeding the current state-of-the-art methods. The significant performance improvement is attributed to block-wise approximation but full-image removal of blocking artifacts. Specifically, with our multi-channel structure, the image blocks with a variety of sampling rates can be reconstructed in a single model. The initially reconstructed blocks are then capable of being reassembled into a full image to improve the recovered images by unrolling a hand-designed block-based CS recovery algorithm. Experimental results demonstrate that the proposed method outperforms the state-of-the-art CS methods by a large margin in terms of objective metrics and subjective visual image quality. Our source codes are available at https://github.com/siwangzhou/DeepBCS.
\end{abstract}

%
\IEEEpeerreviewmaketitle

\section{Introduction}
\label{sec:intro}

Compressive sensing (CS), an emerging sampling and reconstructing strategy, can recover original signal from dramatically fewer measurements with a sub-Nyquist sampling rate \cite{CS,elad2007optimized,CS-MM1}. As CS has the potentials of significantly improving the sampling speed and sensor energy efficiency, it has been applied in many practical applications, including single pixel imaging \cite{singlecamera}, fast magnetic resonance imaging \cite{MRI}, image transmission \cite{Trans-MM-1, Trans-MM-2}, and image encryption \cite{cqli:meet:JISA19}. To deal with high-dimensional natural images efficiently, block-based CS (BCS) is proposed as a lightweight CS approach \cite{BCS,SPL,BCS-1}. In such strategy, a scene under view is partitioned into some small blocks, which are then sampled and reconstructed independently for decreasing the complexity of computation and storage. Moreover, the block partition also benefits adaptive allocation of the sensing resources, since meaningful information is usually not uniformly distributed in an image \cite{BCS-Salie}.

Although block-based CS enjoys the advantages of low-cost sampling, lightweight reconstruction, and capability of adaptively assigning sensing resources, it also usually suffers from reduced quality of image reconstruction due to blocking artifacts \cite{BCS-1,artifact}. To address this issue, the iterative smoothed projected Landweber (SPL) algorithms are proposed \cite{BCS,BCS-1}. In each iteration, the projection operation is used to build an approximation of each block, while smoothing operation acts on the full image reassembled by the approximative blocks. Results demonstrate that the recovered image blocks can be improved, while blocking artifacts can also be ameliorated as the iterations progress. This approach, however, may increase the reconstructing time, since small blocks still need to be concatenated into large-size full images to remove blocking artifacts.

\IEEEpubidadjcol 

Inspired by the powerful learning ability of deep neural networks in image representation \cite{DNN-C, galteri2019deep}, several network-based CS methods are proposed \cite{Reconnet,Ldamp,Im-recon,ISTA}, which are significantly faster than the hand-crafted CS reconstruction algorithms. Using a fully connected layer to mimic the CS sampling, the network models can jointly optimize the sampling matrix and the reconstruction process, improving the qualities of recovered images. Although these network models are carefully constructed to enhance learning capabilities, a CS procedure has to be mapped by a specific model. As a result, when employed in block-based CS, they often fail to take into account the mutual relationships among blocks within an image, especially when multiple sampling rates are assigned to different blocks within an image. Blocking artifacts is then produced \cite{Reconnet,Im-recon,ISTA}. Moreover, most network-based image CS methods are trained as a black box, ignoring structural insight of CS reconstruction algorithms. Consequently, the reconstruction accuracy is decreased.

In this paper, we propose a multi-channel deep learning structure for block-based image CS, termed as BCSnet, which consists of a channel-specific sampling network and a unified deep reconstruction network. The channel-specific structure is specifically designed to handle block-wise sampling rate allocation. The blocks with various sampling rates are then fed into the unified deep reconstruction network to exploit inter-block correlation. We further divide the reconstruction network into a fixed number of residual layers, with which the traditional hand-crafted BCS reconstruction algorithm can be casted into the learning network. The proposed BCSnet is then capable of benefiting from the speed and learning capacities of deep networks while retaining the advantages of the hand-designed BCS algorithm. To enable training, a modified version of the famous DnCNN network designed in \cite{DnCNN} is used to replace the traditional smoothing operation in the SPL approach, which easily propagates gradients, and fortunately offers improved performance.

Our contributions of the paper are summarized below:
\begin{itemize}
\item A multi-channel sampling structure specifically for BCS is designed. Using this multi-channel structure, block-wise CS measuring processed with a variety of sampling rates can be integrated into a single model to utilize the correlation among the blocks.

\item A deep reconstruction structure for block-wise sampling is proposed using block-wise approximation and full-image-based denoising. The merits of the hand-designed BCS reconstruction algorithm can then be combined with the speed and learning capacities of network-based CS approaches, removing the blocking artifacts and improving the recovered images.

\item Performances of the proposed approaches are verified by extensive experiments on the widely used benchmark datasets. The results show that the proposed multi-channel deep network can significantly outperform the state-of-the-art CS methods and network-based ones in terms of both subjective and objective metrics.
\end{itemize}

The remainder of this paper is organized as follows. In Sec.~\ref{sec:rela}, the related work on CS methods and network-based methods are reviewed. Section~\ref{sec:bcsdamp} introduces the idea of block-wise approximation and full-image-based smoothing, and presents BCS-DAMP algorithm. The proposed multi-channel deep structure is presented in Sec.~\ref{sec:dnn} and test results on its performance are given in Sec.~\ref{sec:simu}. The last section concludes the paper.

\section{The related work}
\label{sec:rela}

In this section, we present the background of CS theory, then review the representative work on block-based image CS and deep network approaches.

\subsection{Preliminary of CS theory}
\label{sec:cs}

A CS procedure consists of two main steps: sampling process (measuring process) and reconstructing process. Let $\vct{x}$ and $\Fee$ denote a sparse signal of size $n\times 1$ and an $m \times n$ sampling matrix (measurement matrix), respectively.
Then, the sampling process can be presented as
\begin{equation}
\label{eq:cs}
\vct{y}=\Fee \vct{x},
\end{equation}
where $\vct{y}$ is the $m$-length measurement vector sampled from $\vct{x}$. If signal $\vct{x}$ is not sparse but compressible or approximately sparse in some transform domain, the sampling process has to be further deduced from Eq.~(\ref{eq:cs}). That is, $\vct{y}=\Fee \vct{x} = \Fee \cdot \psi \alpha = \phi \alpha_0 + \upsilon$, where $\upsilon$ represents measurement noise, $\psi$ denotes the sparse transform, $\alpha$ is the coefficient vector in the transform domain, $\phi=\Fee \psi$, and $\alpha_0$ is considered as a sparse vector approximating to $\alpha$. In general, a natural image is not strictly sparse signal, but often compressible.

The reconstructing process of signal $\vct{x}$ needs much more computational complexity than the sampling process. It has been proven in \cite{CS} that if the sampling matrix obeys the restricted isometry property (RIP), it is possible to recover $\vct{x}$ by solving an $l_1$-norm optimization problem: $\min \|\alpha\|_1$, subjecting to $\pnorm{2}{\vct{y}-\phi\alpha}^2 \le \lambda$, even if $m\ll n$, where $\lambda$ is a small constant. One has $\vct{x}=\psi \alpha$ when $\vct{x}$ is a compressible signal.

In the past two decades, a number of CS reconstruction algorithms have been developed, including basis pursuit \cite{BP}, matching pursuit \cite{OMP}, group-based sparse representation GSR algorithm \cite{GSR}, nonlocal low-rank regularization algorithm \cite{dong2014compressive}, and the latest iteration-based DAMP algorithm \cite{Damp}. The authors in \cite{zhu2018on} propose a collaborative CS approach to share the same sampling matrix. Multi-level residual approach and multiple regulation constraints in the recovery process are exploited in \cite{MTAP, chen2017compressive}. Those traditional CS algorithms enjoy solid mathematical foundations, however, they usually need long reconstructing time due to high computational complexity.

\subsection{Block-based image CS}
\label{sec:bcs}

Block-based image CS is a lightweight CS approach, which is more effective for processing natural images because of increased dimensionality of such signals \cite{BCS, BCS-1, BCS-Small}. The scene under view is partitioned into relatively smaller non-overlapping blocks. The measurement matrixes corresponding to the small-size blocks are observed. Then, the image is sampled and reconstructed on a block-by-block basis \cite{BCS-1}. This block-independent approach results in a much simplified sampling process with a reduced computational complexity for reconstruction.

\begin{figure}[!htb]
\centering
\includegraphics[width=0.4\OneImW]{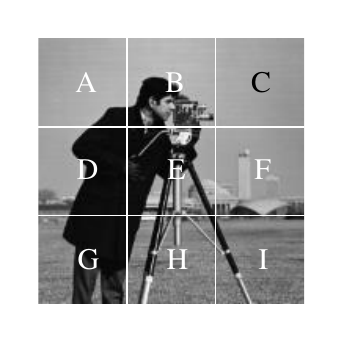}
\includegraphics[width=0.4\OneImW]{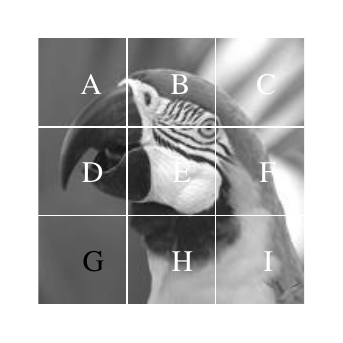}
\caption{Images with block partition.}
\label{fig:block}
\end{figure}

Another advantage of block-based CS is that multiple sampling rates can be easily assigned to different regions within an image. As shown in Fig.~\ref{fig:block}, the meaningful information in different blocks of an image is non-uniform. Lower CS sampling rates should be allocated into the block marked ``C" in the image ``Cameraman" and block ``G" in image ``Parrot", since the information contained in these two blocks is obviously low volume. The adaptive CS sampling techniques based on block partition are recently exploited \cite{BCS-Salie,adaptive,video}. The authors in \cite{BCS-Salie} propose to first acquire an initially sampled image of the scene with a low-resolution complementary optical sensor, and those in \cite{adaptive} propose to start the sensing process with a low fixed-rate sampling and an initial recovery at the receiver. Using the initially sampled or recovered image, the visual significance of the scene is estimated with the visual saliency detection algorithm, and the different sampling rates of the blocks can then be decided before the regular sensing procedures. For compressive video sensing investigated in \cite{video}, the visual significance map can be inferred from previous frame by the use of the correlation between consecutive frames, and it is then easier to achieve the allocation of the sampling rates. In \cite{Asymmetric}, we also propose an asymmetric approach to ensure fairer allocation of sampling rates among image blocks.

However, due to the fact that block partition breaks the global correlation of the whole image, block-wise CS sampling is prone to generate reconstructed image with low quality. In \cite{BCS-Small, BCS-1}, the iterative BCS-SPL reconstruction approach is proposed to remove the blocking artifacts. Let $\hat{X}^{0}=\{\hat{\vct{x}}_i^{0}\}_{i=1}^K$ denote the initial approximation of the whole image, $K$ is the number of image blocks. Initializing index $t=0$, BCS-SPL algorithm is described as follows:
\begin{itemize}
  \item \textit{Approximating}:
  For the $i$-th block $\hat{\vct{x}}_i$ ($1\le i\le K$), one calculates its $(t+1)$-th approximation, $\vct{r}_i^{t+1}$, by employing projection onto convex set,
  \begin{equation}
    \vct{r}_i^{t+1}=\hat{\vct{x}}_i^{t}+\Fee_{\rm B}^*(\vct{y}_i-\Fee_{\rm B} \hat{\vct{x}}_i^{t}),
    \label{eq:spl}
  \end{equation}
  where $\Fee_{\rm B}$ is the measurement matrix corresponding to the block and $\Fee_{\rm B}^*$ is its pseudo-inverse. In the special case when $\Fee_{\rm B}$ is an orthogonal matrix, the pseudo-inverse of $\Fee_{\rm B}$ is equal to its transpose, i.e., $\Fee_{\rm B}^* = \Fee_{\rm B}^T$.

  \item \textit{Hard thresholding smoothing}:
  Reassemble $\{\vct{r}_i^{t+1}\}_{i=1}^K$ into a full image $R^{t+1}$, and calculate
  \begin{equation}
    \hat{X}^{t+1}=\mathcal{H}(R^{t+1}),
    \label{eq:spl22}
  \end{equation}
  where $\mathcal{H}(\cdot)$ represents the hard thresholding function.

  \item \textit{Repetition}:
  Set $t=t+1$ and repeat the above two steps till the difference
  between $\hat{X}^{t+1}$ and $\hat{X}^{t}$ is less than a given threshold.
  \end{itemize}

The recovered images via BCS-SPL algorithm are approximated on a block-by-block basis, but hard-thresholding smoothing in each iteration is imposed on the full image, not image blocks. As a consequence, the artifacts incurred by block partition can be ameliorated as the iteration progresses, while the recovered images are improved by the approximation strategy. Unfortunately, this results in substantially increasing reconstruction complexity because of full-image denoising, which violates the motivation of lightweight design.

\subsection{Deep network approach for image CS}

The tremendous success of deep learning in computer vision attracted application of deep neural networks in image CS \cite{Ldamp,ISTA,Fullnet,Reconnet}. When an imaging system acquires CS measurements, the reconstructing process is performed with a deep network. Compared with the traditional CS, deep network approaches generally enjoy much faster reconstruction speed, while still achieving high-quality recovered images owing to their significant learning capabilities.

In \cite{Reconnet}, the network-based ReconNet approach is introduced to learn the inverse mapping from block-wise CS measurements to their desired image blocks. It is further improved in \cite{Im-recon} that the measurement matrix and the reconstruction process are jointly learned. Their approaches employ traditional off-the-shelf denoiser to remove blocking artifact, but the benefit is not convincingly demonstrated. Compared with the ReconNet approach, Ista in \cite{ISTA} can suit the full-image CS procedure, but it is actually used for block partition due to the high complexity of full-image initial reconstruction and Soft processing. Hence the block artifacts are still be observed, especially when the very low sampling rates are employed. The authors in \cite{Fullnet} and \cite{Csnet} propose to acquire measurements block-by-block but use all measurements from every block to recover the full image, achieving good recovery performance. The performance improvement is due to the fact that full-image convolution is employed to ameliorate the blocking effect. However, they do not consider how to utilize the approximation operation proved effective in the popular BCS-SPL algorithm, and their approaches can still be improved. Moreover, their approaches can not apply to the scenarios where different sampling rates are assigned to the blocks, according to their network architecture.

In general, a CS procedure with a specific sampling rate is mapped by a separate network model. That is, a number of network models have to be employed to deal with an image with the adaptive block-wise CS sampling. As a result, too many model parameters need to be stored beyond the blocking artifacts. We notice that a neural network structure illustrated in \cite{Ldamp} is applied to a variety of measurement matrices. Unfortunately, its performance improvement is less significant than traditional CS methods because the measurement matrices cannot participate in the network training. Inspired by the multi-scale super-resolution method given in \cite{Enhanced}, the authors in \cite{Multi} introduce a multi-scale CS approach, where the main network is shared across multiple sampling rates. However, their method only reuses a portion of parameters, and a CS sampling rate still corresponds to a specific network model. In other words, they do not consider block partition and the corresponding problem of blocking artifacts.

\section{BCS-DAMP algorithm}
\label{sec:bcsdamp}

In this section, we propose an extension for DAMP algorithm, called BCS-DAMP, aiming at providing a link between the traditional BCS-SPL algorithm and our deep network based solution in the following section.

DAMP is a state-of-the-art CS reconstruction algorithm, which is also an iterative approach like BCS-SPL. DAMP approach proposes to employ a generic denoiser to replace the smoothing operation in general SPL recovery algorithm, and further demonstrates that, when used with a high-performance denoiser the state-of-the-art recovery performance can be offered. However, DAMP is not specially designed for block-based CS, and does not consider the blocking artifacts at all.

We then propose an extended version of DAMP, BCS-DAMP, for block-based image CS:

\begin{itemize}
  \item \textit{Estimate of the residual}:
  For the $i$-th block $\hat{\vct{x}}_i^t$ ($1\le i\le K$) at $t$-th iteration, one estimates its residual $\vct{z}_i^t$,
  \begin{equation}
   \vct{z}_i^t=\vct{y_i}-\Fee_{B}\hat{\vct{x}}_i^t+\vct{z}_i^{t-1} \{{\rm div} \mathcal{D}_{\hat{\sigma}^{t-1}}(R^t)\}_i/m,
   \label{eq:damp-esti}
  \end{equation}	
  where ${\rm div}$ denotes the operation of partial derivative, ${\rm div}\mathcal{D}_{\hat{\sigma}^{t-1}}$ represents the divergence of the denoiser, $m$ is the number of CS measurements, $\{\cdot\}_i$ represents the $i$-th block of the object, $R^t$ indicates the full images concatenating of all $K$ recovered blocks at $t$-th iterations. $\vct{z}_i^0$ is initialized to $\vct{y}_i$.

  \item \textit{Approximating}:
  For block $\hat{\vct{x}}_i^t$, one calculates its $(t+1)$-th approximation, $\vct{r}_i^{t+1}$:
  \begin{equation}
    \vct{r}_i^{t+1}=\hat{\vct{x}}_i^{t}+\Fee_{\rm B}^*\vct{z}_i^t.
    \label{eq:damp-appro}
  \end{equation}

  \item \textit{Denoising}:
  \begin{equation}
    \hat{\vct{x}}_i^{t+1} = \mathcal{D}_{\hat{\sigma}^t}(R^{t+1}),
    \label{eq:damp-deno}
  \end{equation}
  where $\mathcal{D}_{\hat{\sigma}^t}(\cdot)$ represents a general denoising function, and $\hat{\sigma}^t =\parallel \vct{z}_i^t \parallel_2/\sqrt{m}$, which is an estimate of the standard deviation of that noise.

  \item \textit{Repetition}:
  Set $t=t+1$ and repeat the above two steps till the difference
  between $\hat{\vct{x}}_i^{t+1}$ and $\hat{\vct{x}}_i^t$ satisfies a given condition.
\end{itemize}

The key enhancement to the DAMP algorithm is that, $\mathcal{D}_{\hat{\sigma}^t}(\cdot)$ does not run on the blocks, but run on the full image obtained by concatenating all approximated blocks. In this way, our BCS-DAMP enjoys both the benefit of full-image processing in BCS-SPL and the advantage of high-performance denoising in DAMP.

We think more about the full-image denoising in the BCS-DAMP, and hold that, if higher-performance network based denoising strategy is employed, the CS recovery performance may be further improved. In the following section, network based full-image denoising and block-wise iterative approximation will be casted into a carefully designed deep network for removing artifacts and improving the recovered images.

\section{Multi-channel deep network structure}
\label{sec:dnn}

\begin{figure}[!htb]
\centering
\includegraphics[width=1.1\OneImW]{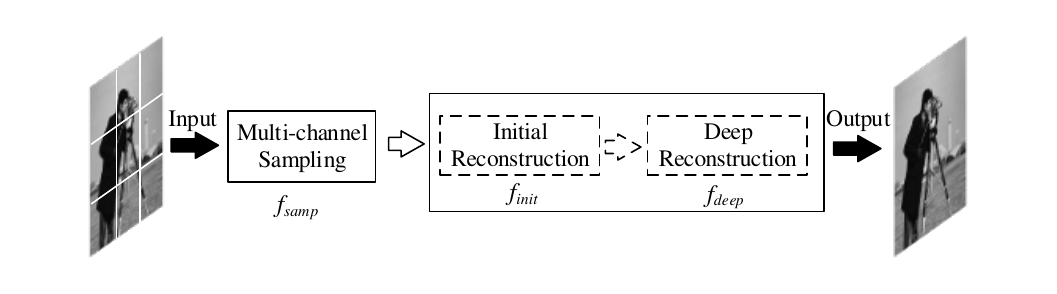}
\caption{The structure of BCSnet.}
\label{fig:framework}
\end{figure}

In this section, we propose a multi-channel deep network structure to reconstruct the images acquired by block-wise CS sampling. To facilitate description, we term it as BCSnet. As shown in Fig.~\ref{fig:framework}, our BCSnet is composed of a multi-channel sampling network and a deep reconstructing network. These two networks consist of an integrated end-to-end model, whose parameters are jointly trained by the proposed two-stage training strategy.

\subsection{The structure of multi-channel sampling}
\label{sec:samp}

In general block-based image CS algorithms, fixed-rate allocation strategy is employed in the sampling process, where all blocks in the image are assigned the same sampling rate. Adaptive sampling strategies introduced in \cite{BCS-Salie,adaptive} propose to assign different sampling rates to different blocks, since the meaningful information is usually not uniformly distributed in an image.

\begin{figure}[!htb]
\centering
\includegraphics[width=\OneImW]{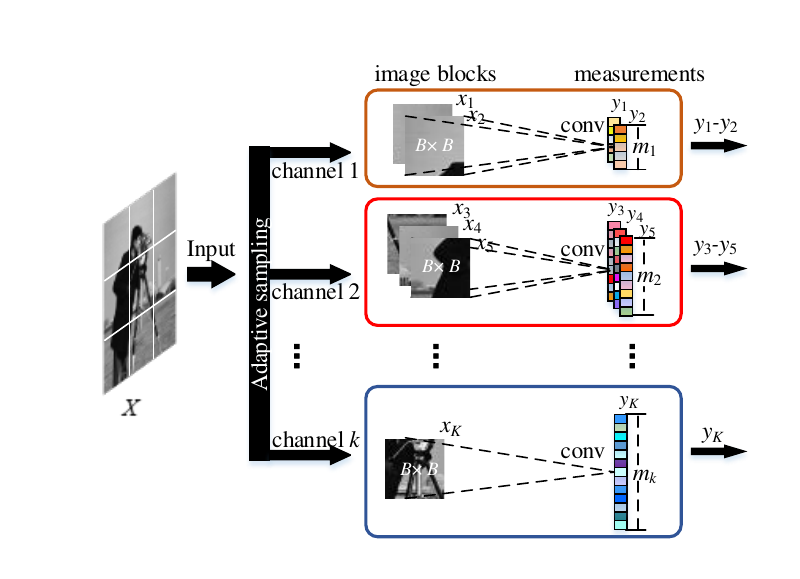}
\caption{The structure of $k$-channel sampling network $f_{\rm samp}$.}
\label{fig:sample}
\end{figure}

This subsection investigates a $k$-channel network structure, named $f_{\rm samp}$, to mimic the adaptive sampling process of block-based CS, as shown in Fig.~\ref{fig:sample}. Note that our multi-channel structure also fits into the fixed-rate allocation strategy of CS sampling.

Network $f_{\rm samp}$ has $k$ channels, each of which corresponds to a specific sampling rate. A higher value of $k$ indicates a more detailed division of sampling rates for image blocks. Block partition can benefit more fairer allocation of the sensing resources, but resulting in more complex sampling structure. Note that how to get the value of $k$, or the number of sampling rates, is a practical issue. One can make a decision based upon empirical observations, since the training images are easily acquired and analysed ahead of the sensing procedure.

For the $j$-th channel in $f_{\rm samp}$, a convolutional layer without bias and activation is used to mimic the sampling operation. The convolutional layer is defined as $w_j^{\rm samp}\otimes \vct{x}_i$, where $w_j^{\rm samp}$ denotes $m_j$ convolution kernels of size $B \times B \times 1$. In other words, $w_j^{\rm samp}$ corresponds to measurement matrix $\Fee_{{\rm B},j}$ in CS sampling operation, where the size of an image block is $B \times B$ and the number of measurements is $m_j$. If $\vct{x}_i$ is fed into the $j$-th channel, one has
\begin{equation}
  \vct{y}_i=f_{\rm samp}(\vct{x}_i, w_j^{\rm samp}).
  \label{eq:samp}
\end{equation}
At this point, $k$ sets of convolutional kernels, $\{w_j^{\rm samp}\}_{j=1,2,\cdots,k}$, have to be stored, and this imposes additional memory consumption. Fortunately, the memory consumption can be mitigated by compacting the corresponding convolutional kernels. The interested readers can refer to \cite{cheng2018model}.

For the proposed multi-channel network, a channel is related to an interval of sampling rates. In the training procedure, the sampling rate for each block of the images in the training dataset is pre-computed according to its visual saliency. The block then enters the sampling network by selecting a channel according to its sampling rate. The measurements among different channels are different from each other, since each channel is related to its own convolutional kernel. However, the blocks within the same image are spatially correlated. As a consequence, the measurements of those blocks among different channels are related to each other. Our multi-channel sampling structure makes this inter-block correlation available in the deep reconstruction network.

We should note that, the proposed multi-channel structure is specially designed to mimic the adaptive allocation strategy, where multiple sampling rates may be assigned to the blocks within an image. However, our scheme also fits the scenario where all blocks within an image are with the same sampling rates. For the application scenario where each block is assigned the same sampling rate, all blocks can be input the network via the same channel. At this point, our $k$-channel sampling model degrades into a general sampling network, as employed by the existing network-based CS methods. In contrast, our multi-channel scheme works better since it can deal with $k$ sampling rates within one network.

\subsection{The structure of deep reconstruction}

\begin{figure*}[!htb]
	\centering
	\includegraphics[width=1.6\OneImW]{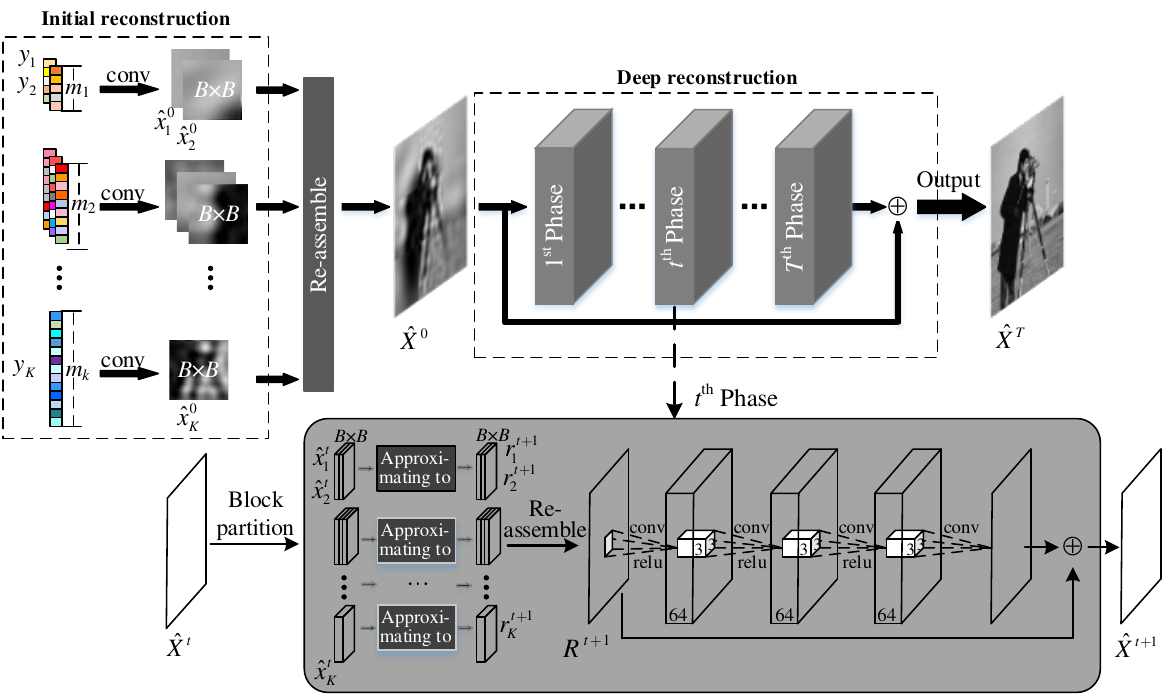}
	\caption{The structure of deep reconstruction consisting of initial reconstruction $f_{init}$ and deep reconstruction $f_{deep}$.}
	\label{fig:recon}
\end{figure*}

\begin{figure}[!htb]
\centering
\includegraphics[width=0.8\OneImW]{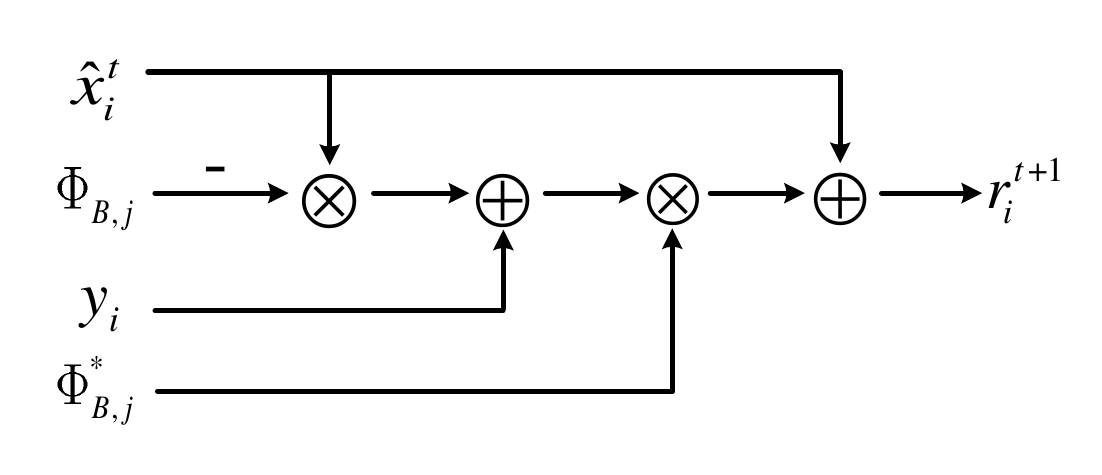}
\caption{The feed-forward network implementation of the approximating operation.}
\label{fig:approx}
\end{figure}

In this subsection, we construct a deep reconstruction structure to cast the idea of iterative block-wise approximation but full-image-based denoising into the network, improving the recovered blocks while achieving removal of blocking artifacts from the level of model. The proposed scheme is composed of an initial reconstruction network $f_{\rm init}$ and a deep reconstruction network $f_{\rm deep}$, as shown in Fig.~\ref{fig:recon}. In deep reconstruction network, the approximating operation at $t$-th phase is implemented by using a feed-forward network, as illustrated in Fig.~\ref{fig:approx}. The approximated blocks are reassembled into a full image to perform denoising by employing a modified version of DnCNN phase by phase.

The initial reconstruction network, $f_{\rm init}$, has $k$ inputs, each of which corresponds to a sampling channel in $f_{\rm samp}$ as illustrated in Sec.~\ref{sec:samp}. The $j$-th input is connected to the corresponding $j$-th convolutional layer, which uses $B^2$ kernels of size $1\times 1 \times m_j$ and generates $B^2$ values by convolving them with $\vct{y}_i$, the measurement of block $\vct{x}_i$ entering the $j$-th channel. All the $B^2$ values are combined into one $B \times B$ feature map, $\hat{\vct{x}}_i^0$, which is used as the initially reconstructed result of block $\vct{x}_i$. From the view point of network, one has
\begin{equation}
\hat{\vct{x}}_i^{0}=f_{\rm init}(\vct{y}_i, w_j^{\rm init}),
\end{equation}
where $w_j^{\rm init}$ denotes the above mentioned $1\times 1 \times m_j$ convolutional kernel corresponding to the $j$-th channel in the sampling network. As shown in Fig.~\ref{fig:recon}, the initial reconstruction network includes only one convolutional layer for simplification reason, and the initially recovered images are improved further by the deep reconstruction network.

The proposed deep reconstruction network, $f_{\rm deep}$, is further divided into $T$ phases, so that the iterative BCS-SPL algorithm can be unrolled along with the $T$ phases. In $f_{\rm deep}$, each phase corresponds to one iteration in BCS-SPL algorithm consisting of approximating and denoising operations. In the $(t+1)$-th phase, the block-wise approximation is implemented by using a formula slightly different from Eq.~(\ref{eq:spl}) shown in Sec. \ref{sec:bcs}.
It becomes
\begin{equation}
  \label{eq:proj-spl}
  r_i^{t+1}=\hat{\vct{x}}_i^{t}+\Fee_{{\rm B},j}^*(y_i-\Fee_{{\rm B},j}\hat{\vct{x}}_i^{t})
\end{equation}
for each block $\vct{x}_i$, where $\Fee_{{\rm B},j}$ is the measurement matrix specialized for block $\vct{x}_i$, if $\vct{x}_i$ is fed into the network via $j$-th channel in the multi-channel sampling model. Note that the matrix $\Fee_{{\rm B},j}$ is learnable, and it may be not an orthogonal matrix. Thus, its pseudo-inverse, $\Fee_{{\rm B},j}^*$, can not be simplified into $\Fee^T_{{\rm B},j}$ as in the traditional BCS-SPL algorithm.

All approximated blocks $\{r_i^{t+1}\}_{i=1}^K$ are reassembled to build an approximate version of the whole image, $R^{t+1}$, for further denoising processing. To enable training of the deep reconstruction network, the famous DnCNN network is modified to implement full-image denoising. Traditional denoising methods, such as hard thresholding in BCS-SPL and BM3D in DAMP algorithm, do not work in deep network structure, since they cannot propagate gradients. This restricts focus on feed-forward convolutional neural networks. Fortunately, DnCNN offers improved performance on image deblocking and Gaussian denoising. The proposed reconstruction network is composed of $T$ phases. Each phase has $l$ convolutional layers, and the configuration is designed by referring to the DnCNN network. The first layer generates $d$ feature maps with the $d$ kernels of size $f\times f \times 1$, the following two layers employ $d$ kernels, each of which is of size $f\times f\times d$, and the last layer generates one feature map with an $f \times f \times d$ kernel. It should be noted that all conventional layers explore the relu activation function except the last layer. In DnCNN, 20 convolutional layers are employed to form a deep network for image denoising, while in our network model, $l$ convolutional layers form a phase and $T$ phases are employed to deal with both image denoising and image approximation. In the experiments, $d$, $f$, $l$, and $T$ are set to 64, 3, 4, and 10, respectively. Let $w_t^{\rm deep}$ be the parameters of convolutional kernels in $t$-th phase. Then one has
\begin{equation}
  \hat{X}^{T}=f_{\rm deep}(R^{1}, \{w_t^{\rm deep}\}_{t=1}^T),
  \label{eq:phase}
\end{equation}
where $R^1$ is the approximated concatenated image in the first phase of the deep network.

\subsection{Two-stage training}

As illustrated in Sec.~\ref{sec:samp}, sampling matrixes $\{\Fee_{{\rm B},j}\}_{j=1}^k$ in the network are implemented by employing convolution operations. That is, the elements in $\Fee_{{\rm B},j}$ and $\Fee_{{\rm B},j}^*$ have to be taken from the convolution kernel in the $j$-th channel in the sampling network. However, if both $\Fee_{{\rm B},j}$ and $\Fee_{{\rm B},j}^*$ participate in the training process, a desired recovered image can not be obtained. This is caused by that $\Fee_{{\rm B},j}$ has to be updated in real time along with each back propagation in the training process, and back propagation is based on the gradient decent rule, which is hindered due to the real-time computing of $\Fee_{{\rm B},j}^*$ in the deep reconstruction network.

To improve the recovered images by training sampling matrix $\{\Fee_{{\rm B},j}\}_{j=1}^k$ while utilizing $\{\Fee_{{\rm B},j}^*\}_{j=1}^k$ in deep reconstruction process, the training process is divided into two stages.

The first stage aims to obtain the training parameters of the sampling network. That is, we first achieve $k$ optimal sampling matrices $\{\Fee_{{\rm B},j}\}_{j=1}^k$ and the corresponding matrices $\{\Fee_{{\rm B},j}^*\}_{j=1}^k$.
The proposed deep structure consists of an initial reconstruction network $f_{\rm init}$ without including $\Fee_{{\rm B},j}^*$ and a deep reconstruction network $f_{\rm deep}$, where $\Fee_{{\rm B},j}^*$ is utilized to improve the recovered images. In this way, the sampling network and the initial reconstruction part of our reconstruction network are combined into a training network, i.e., $f_{\rm init}(f_{\rm samp}(\cdot), \cdot)$, which is used to train the sampling matrix $\{\Fee_{{\rm B},j}\}_{j=1}^k$. Given the training images $\{X_n\}_{n=1}^N$, the loss function of the first-stage training network is:
\begin{multline}
L_{\rm samp}=\frac{1}{2N}\sum_{n=1}^N \Vert f_{\rm init}(f_{\rm samp}(X_n, \{w_j^{\rm samp}\}_{j=1}^k), \\
\{w_j^{\rm init}\}_{j=1}^k)-X_n \Vert^2,
\label{eq:loss-fir}
\end{multline}
where $N$ is the number of images in the training dataset.

In the second stage, we further train the reconstructing network consisting of an initial reconstruction part and a deep reconstruction part, i.e., $f_{\rm deep}(f_{\rm init}(\cdot), \cdot)$, where the parameters in $\{\Fee_{{\rm B},j}^*\}_{j=1}^k$ come from $\{\Fee_{{\rm B},j}\}_{j=1}^k$. That is, the sampling weights $\{w_j^{\rm samp}\}_{j=1}^k$ are fixed while the parameters $\{\{w_j^{\rm init}\}_{j=1}^k, \{w_t^{\rm deep}\}_{t=1}^T\}$ are updated in the training process. Reconstructing network $f_{\rm deep}(f_{\rm init}(\cdot), \cdot)$ directly learns the mapping between the CS measurements and the ground truth, and the loss function minimizing the error between the input and the output relies on the full images instead of image blocks. Mean square error is adopted to design an end-to-end loss function
\begin{multline}
L_{\rm deep}=\frac{1}{2N}\sum_{n=1}^N \parallel f_{\rm deep}(f_{\rm init}(\{\vct{y}_{i,n}\}_{i=1}^K, \\
  \{w_j^{\rm init}\}_{j=1}^k), \{w_t^{\rm deep}\}_{t=1}^T)-X_n \parallel_2^2,
  \label{eq:loss}
\end{multline}
where $\vct{y}_{i,n}$ denotes the CS measurement vector of the $i$-th block in the $n$-th training image.

\section{Performance evaluation}
\label{sec:simu}

In this section, we perform extensive experiments to evaluate the performance of the proposed BCSnet and BCS-DAMP schemes, and compare them with state-of-the-art methods, including video CS models 2-D TV and 3-D TV \cite{video}, traditional BCS-SPL \cite{BCS-1}, DAMP \cite{Damp}, GSR \cite{GSR}, network-based Ista \cite{ISTA}, ReconNet \cite{Reconnet} and its improved version, I-Recon \cite{Im-recon} in terms of reconstruction quality, time complexity and visual effect. As for Ista, DAMP, GSR, and BCS-SPL approaches, the authors provide the source codes, and we recompile them in our experimental environment. As of ReconNet and I-ReconNet, we redo the programming implementation according to the network architecture they describe. In the experiments, BM3D presented in \cite{BM3D} is used as the denoiser of DAMP and our proposed BCS-DAMP algorithm.

For objective evaluation, \textit{PSNR} (Peak Signal-to-Noise Ratios), \textit{SSIM} (Structural Similarity Index Measure), and \textit{Perceptual Similarity} are used to measure the quality of recovered images. For subjective evaluation, \textit{MOS} (mean opinion score) marked by various volunteers is employed as the measurement criteria. In the experiments, we set the phase number of our BCSnet model to 10, and the number of convolutional layers for each phase is set to 4. That is, the BCSnet has a total 40 convolutional layers, which is comparative to the famous Ista method. We use TensorFlow 1.4 in \cite{Tensor} to train the proposed multi-channel network at a desktop platform configured with one GPU NVIDIA 1060, one CPU Intel Core i7-4790K with frequency 4.00 GHz and 32GB of memory. The training processes takes about 3 hours for one epoch, and we train our network with 50 epoches.

\subsection{Training and testing}
\label{sec:tr_te}

The parts on training details and test set selection are separately described as follows.

\subsubsection{Training detail}
\label{sec:trainset}

The training images are selected from the training set (200 images) and test set (200 images) of the BSDS500 database as \cite{BSD500}. We randomly crop 89600 images of size $96 \times 96$ as the training set. Each training image, $X$, is further partitioned into nine image blocks of size $32 \times 32$, $\{\vct{x}_{i}\}_{i=1}^9$. So there are a total of 806,400 blocks in our training set. Visual saliency of the scene exploited in \cite{BCS-Salie} is employed for adaptive sampling rate allocation of the training images. Suppose that $v$ represents the amount of the saliency information embodied in image $X$. One has $v=\frac{1}{n} \sum\nolimits_{j\in X_s}l_j$, where $n$ is the total number of pixels on image $X$, $X_s$ denotes the saliency map of $X$, and $l_j$ is the saliency value of location $j$ on $X_s$. Let $v_{i}$ be the saliency information of image block $\vct{x}_{i}$, and $p_{i} = \frac{v_{i}}{v}$ denotes the proportion of the saliency information of block $\vct{x}_{i}$. The used training data pair is $(X, \{(\vct{x}_{i}, p_{i})\}_{i=1}^{K})$, where $K=9$ and there are $N= 89,600$ training image pairs in total. Note that in the testing stage, the sizes of test images remain unaffected by the number of blocks of the training images, $K$, thanks to the proposed block-wise multi-channel network architecture.

As for the Ista, ReconNet and I-Recon, the network models are trained with our training set, where the blocks and themselves consists of 806,400 training block pairs, $\{\vct{x}_i, \vct{x}_{i}\}_{i=1}^{806400}$. All the three approaches employ the same adaptive sampling rate allocation mentioned above, and use block-wise CS reconstruction strategy.

With our BCS-Net, we take $k=7$ as an example for the $k$-channel sampling, and the sampling rates fall in set $\{0.01, 0.03, 0.05, 0.1, 0.2, 0.3, 0.4\}$. We should notice that our training process can also apply to any $k$ values. Each image pair $(X, \{(\vct{x}_{i}, p_{i})\}_{i=1}^9)$ is further processed in order to find out the most appropriate channels in the network. For a given target sampling rate, $SR$, we calculate sub-rate $s_{i}$ of block $\vct{x}_{i}$ as $s_{i}=SR \cdot p_{i}$, where $p_{i}$ is defined in Sec.~\ref{sec:trainset}. In the training process, the space of sampling rates is divided into seven intervals, each of which corresponds to a channel. If $s_{i}$ falls within the interval $[T_c, T_{c+1}]$ for the $c$-th channel, block $\vct{x}_{i}$ is pushed into the network via this channel.

\begin{figure}[!htb]
	\centering
	\begin{minipage}[t]{\twofigwidth}
		\centering
		\includegraphics[width=\twofigwidth]{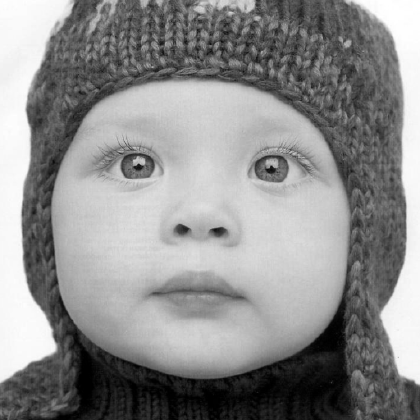}
		a)
	\end{minipage}\enspace
	\begin{minipage}[t]{\twofigwidth}
		\centering
		\includegraphics[width=\twofigwidth]{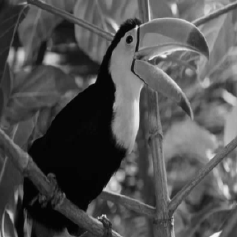}	
		b)
	\end{minipage}
	\begin{minipage}[t]{\twofigwidth}
		\centering
		\includegraphics[width=\twofigwidth]{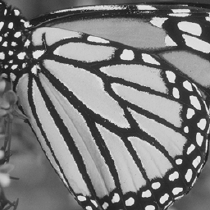}	
		c)
	\end{minipage}
	\begin{minipage}[t]{\twofigwidth}
		\centering
		\includegraphics[width=\twofigwidth]{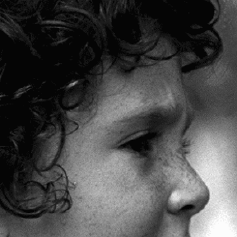}
		d)
	\end{minipage}
	\begin{minipage}[t]{\twofigwidth}
		\centering
		\includegraphics[height=\twofigwidth]{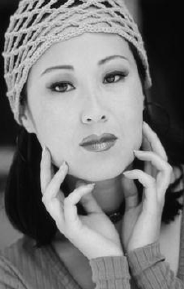}\\	
		e)
	\end{minipage}
	\caption{Five typical images in Set5 owning different spatial information distribution: a) ``Baby"; b) ``Bird"; c) ``Butterfly"; d) ``Head"; e) ``Woman".}
	\label{fig:set5}
\end{figure}

\begin{figure*}[!htb]
	\centering
	\includegraphics[width=2\OneImW]{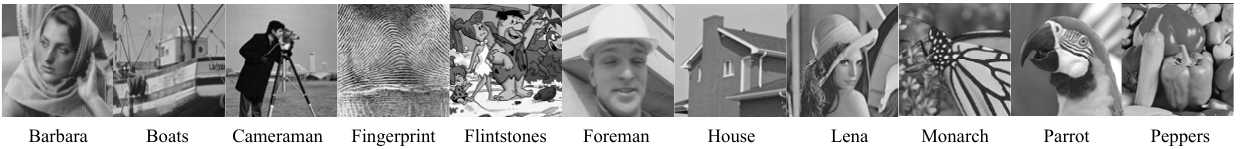}
	\caption{Eleven images with a variety of spatial information distribution in test dataset ``Set11".}
	\label{fig:set11}
\end{figure*}

\subsubsection{Testing}

A 16-frame video and three widely used image sets, Set5, Set11 and BSD100, are used for testing the multi-channel networks. Set5 and Set11 are shown in Fig.~\ref{fig:set5} and Fig.~\ref{fig:set11}, respectively.

The 16-frame video is a talking person ``Foreman" with $288\times 352$ pixel resolution. Set5 consists of 5 gray images, where the sizes of ``Bird" and ``Head" are $288\times 288$, ``Baby", ``Butterfly" and ``Woman" are with the size of $512 \times 512$, $256\times 256$ and $224 \times 352$, respectively. Set11 has 11 gray images, where the sizes of ``Fingerprint" and ``Flintstones" are $512\times 512$, and the other nine images are all of size $256\times 256$. BSD100 includes 100 images, of which the central $320\times320$ parts are extracted as test images. The test images own a various types of spatial distribution of key visual information. For example, the main meaningful information in images ``Cameraman" and ``Parrot" in Set11 is located in the single connected region. In contract, the visual information of ``Bird" in Set5, ``Fingerprint", ``Flintstones" and ``Peppers" in Set11 uniformly distributes in the whole images. Note that all those test images are strictly separate from the training datasets.

In actual applications, the well-trained multi-channel network model is stored in the image processing unit. When the scene under view is sampled with some dedicated optical sensor, the CS measurements are transmitted to this processing unit. The proposed network model can then be employed to recover the original image for testing.

\begin{table*}[!htb]
	\centering
	\caption{The average \textit{PSNR} and \textit{SSIM} on three image sets with adaptive allocation of sampling rates. The best performance scores are labeled in bold.}
	\begin{tabular}{c|ccccccc}
		\hline
		\multicolumn{8}{c}{Set5 (\textit{PSNR}/\textit{SSIM})}\\
		\hline
		\diagbox[width=10em]{Algorithm}{Sampling rate} &0.01 &0.03 &0.05 &0.1 &0.2 &0.3 &0.4\\
		\hline
		BCS-SPL \cite{BCS-1} &16.20/0.3613 &18.62/0.4874 &23.54/0.6451 &25.98/0.7299 &29.06/0.8158 &31.18/0.8618 &32.32/0.8899\\
		DAMP \cite{Damp} &6.51/0.0311 &12.39/0.2981 &21.88/0.5542 &25.07/0.6929 &29.81/0.8298 &33.25/0.9073 &34.25/0.9395\\
		GSR \cite{GSR} &18.88/0.5082 &23.07/0.6783 &25.13/0.7677 &29.56/0.8692 &33.84/0.9297 &36.46/0.9587 &38.53/0.9707\\
		ReconNet \cite{Reconnet} &18.46/0.4492 &22.69/0.5844 &24.51/0.6672 &26.89/0.7518 &29.55/0.8348 &31.20/0.8738 &31.00/0.8793\\
		I-Recon \cite{Im-recon} &21.49/0.5571 &26.31/0.7275 &28.34/0.8022 &30.28/0.8496 &33.12/0.9023 &35.07/0.9357 &35.30/0.9465\\
		Ista \cite{ISTA} &18.06/0.4589 &22.70/0.6210 &25.72/0.7332 &29.91/0.8462 &33.94/0.9154 &36.67/0.9479 &38.14/0.9622\\
		\textbf{BCS-DAMP} &17.69/0.4087 &22.16/0.5909 &24.46/0.7133 &28.28/0.8280 &32.92/0.9083 &35.95/0.9407 &37.42/0.9556\\
		\textbf{BCSnet-WBA} &22.52/0.5844 &26.33/0.7355 &28.65/0.8178 &31.65/0.8853 &35.08/0.9372 &37.54/0.9626 &38.54/0.9723\\
		\textbf{BCSnet-WBA (50 layers)} &22.60/0.5881 &26.51/0.7434 &28.87/0.8254 &31.92/0.8904 &35.29/0.9403 &37.73/0.9641 &38.84/0.9741\\
		\textbf{BCSnet} &\textbf{22.93}/\textbf{0.6037} &\textbf{27.07}/\textbf{0.7655} &\textbf{29.50}/\textbf{0.8420} &\textbf{32.64}/\textbf{0.9010} &\textbf{36.04}/\textbf{0.9462} &\textbf{38.42}/\textbf{0.9665} &\textbf{39.61}/\textbf{0.9762}\\
		\hline
		\multicolumn{8}{c}{Set11 (\textit{PSNR}/\textit{SSIM})}\\
		\hline
		\diagbox[width=10em]{Algorithm}{Sampling rate} &0.01 &0.03 &0.05 &0.1 &0.2 &0.3 &0.4\\
		\hline
		BCS-SPL \cite{BCS-1} &15.65/0.3973 &18.47/0.5120 &21.40/0.5991 &23.38/0.6806 &26.36/0.7806 &28.74/0.8493 &29.75/0.8783\\
		DAMP \cite{Damp} &5.49/0.0582 &9.77/0.2659 &20.36/0.5497 &23.21/0.6722 &27.52/0.8184 &31.32/0.8943 &32.93/0.9339\\
		GSR \cite{GSR} &16.78/0.4549 &19.64/0.5925 &22.76/0.7157 &27.55/0.8527 &32.15/\textbf{0.9338} &34.75/0.9465 &\textbf{36.89}/\textbf{0.9693}\\
		ReconNet \cite{Reconnet} &16.99/0.4145 &20.55/0.5407 &22.09/0.6135 &24.32/0.7079 &26.63/0.7910 &28.24/0.8363 &28.32/0.8411\\
		I-Recon \cite{Im-recon} &19.80/0.5018 &23.72/0.6723 &24.44/0.7527 &27.29/0.8121 &30.15/0.8763 &32.03/0.9137 &32.26/0.9243\\
		Ista \cite{ISTA} &16.55/0.4139 &20.88/0.5767 &23.68/0.6851 &27.64/0.8149 &31.86/0.9018 &34.74/0.9393 &36.03/0.9547\\
		\textbf{BCS-DAMP} &18.13/0.4781 &20.97/0.5815 &22.83/0.6522 &26.17/0.7847 &30.83/0.8974 &34.24/0.9462 &35.60/0.9596\\
		\textbf{BCSnet-WBA} &20.52/0.5241 &23.80/0.6761 &25.75/0.7638 &28.49/0.8463 &31.94/0.9153 &34.43/0.9468 &35.44/0.9583\\
		\textbf{BCSnet-WBA (50 layers)} &20.58/0.5293 &23.96/0.6844 &25.96/0.7715 &28.75/0.8520 &32.16/0.9179 &34.66/0.9483 &35.73/0.9595\\
		\textbf{BCSnet} &\textbf{20.81}/\textbf{0.5427} &\textbf{24.41}/\textbf{0.7041} &\textbf{26.50}/\textbf{0.7893}  &\textbf{29.36}/\textbf{0.8650} &\textbf{32.87}/0.9254 &\textbf{35.40}/\textbf{0.9527} &36.52/0.9640\\
		\hline
		\multicolumn{8}{c}{BSD100 (\textit{PSNR}/\textit{SSIM})}\\
		\hline
		\diagbox[width=10em]{Algorithm}{Sampling rate} &0.01 &0.03 &0.05 &0.1 &0.2 &0.3 &0.4\\
		\hline
		BCS-SPL \cite{BCS-1} &18.53/0.4213 &18.95/0.4476 &22.09/0.5204 &23.70/0.5907 &25.71/0.6852 &27.23/0.7487 &27.75/0.7875\\
		DAMP \cite{Damp} &7.00/0.0495 &10.71/0.1950 &21.00/0.4569 &22.80/0.5312 &25.11/0.6335 &26.80/0.7093 &27.57/0.7541\\
		GSR \cite{GSR} &18.03/0.4227 &19.52/0.4852 &21.06/0.5529 &23.18/0.6635 &26.69/0.7840 &29.13/0.8522 &31.17/0.8966\\
		ReconNet \cite{Reconnet} &18.74/0.3960 &21.29/0.4754 &22.34/0.5253 &23.86/0.5969 &25.51/0.6837 &26.65/0.7400 &26.71/0.7611\\
		I-Recon \cite{Im-recon} &21.15/0.4654 &23.84/0.5767 &24.87/0.6419 &26.30/0.7228 &28.50/0.8121 &29.82/0.8593 &30.07/0.8812\\
		Ista \cite{ISTA} &17.86/0.3957 &21.57/0.5041 &23.14/0.5677 &25.75/0.6822 &28.78/0.8012 &30.87/0.8660 &31.70/0.9003\\
		\textbf{BCS-DAMP} &19.41/0.4199 &21.06/0.4770 &22.47/0.5233 &24.20/0.5849 &26.75/0.6827 &28.97/0.7623 &30.21/0.8241\\
		\textbf{BCSnet-WBA} &21.75/0.4836 &24.07/0.5913 &25.29/0.6593 &27.35/0.7542 &30.03/0.8552 &31.98/0.9064 &32.71/0.9319\\
		\textbf{BCSnet-WBA (50 layers)} &21.79/0.4860 &24.13/0.5945 &25.36/0.6626 &27.41/0.7570 &30.08/0.8569 &32.06/0.9080 &32.81/0.9336\\
		\textbf{BCSnet} &\textbf{21.97}/\textbf{0.4943} &\textbf{24.38}/\textbf{0.6068} &\textbf{25.64}/\textbf{0.6754} &\textbf{27.78}/\textbf{0.7688} &\textbf{30.51}/\textbf{0.8652} &\textbf{32.54}/\textbf{0.9137} &\textbf{33.33}/\textbf{0.9383}\\
		\hline
	\end{tabular}
	\label{tab:compa}
\end{table*}

\subsection{Results and analysis}
\label{sim:result}

\subsubsection{Performance evaluation with adaptive allocation of sampling rates}
\label{sim:compa}

The performances of the proposed BCS-DAMP and BCSnet were evaluated, and compared with the existing methods. BCSnet-WBA, the weaken version of BCSnet without block-wise approximation, is also used for the evaluation.

The sampling rates of the test images are first computed according to the visual saliency of each block. Then each sampling value is rounded by keeping only one valid decimal place, so that it can correspond to one of the sampling rates of $k$ channels, $\{0.01, 0.03, 0.05, 0.1, 0.2, 0.3, 0.4\}$ in this experiment. For example, if the sampling rate of a block calculated from the test image is 0.1136, then the actual rate we use for testing is rounded to 0.1. In a sense, the rough rounding operation can be considered as an estimation of the ground truth, similarly to the initial sampling techniques in \cite{BCS-Salie, adaptive}. The different sampling rates of the blocks are also used by the competing methods, including network-based ReconNet, I-Recon, Ista, and traditional BCS-SPL, DAMP, for a fair comparison. But as for GSR algorithm, due to excessive amounts of running time to reconstruct an image, the reconstruction experiment is not with the adaptive CS sampling, but with the simple fixed-rate allocation.

The average quality scores of the output images of BCSnet in terms of \textit{PSNR} and \textit{SSIM} are given in Table~\ref{tab:compa}. The detail scores of the output images on Set5 are given in Table~\ref{tab:set5psnr}. Observing Table~\ref{tab:compa}, one can see that BCSnet yields higher-quality recovered image than other methods, including BCS-SPL, DAMP, GSR, ReconNet, I-Recon, and Ista. The significant performance improvement between our BCSnet and the best results of the existing methods can be observed: 2.36(32.64-30.28) dB on Set5, 2.06(26.50-24.44) dB on Set11, and 1.73(30.51-28.78) dB on BSD100 at the sampling rate of 0.1, 0.05, and 0.2, respectively. The performance improvement of the proposed BCSnet is mainly attributed to the following two factors: block-wise approximation and full-image denoising. This can be validated in a way by our BCSnet-WBA and that with 50 layers, the weaken versions of BCSnet, in which full-image denoising is employed but block-wise approximation is not used. As shown by the quality scores of BCSnet-WBA (50 layers), when the depth of our network increases from 40 to 50 convolutional layers, the recovered images are not improved much. This indicates that the performance improvement mainly comes from the strategy of block-wise approximation, not the deeper network. From Table~\ref{tab:compa}, our BCS-DAMP achieves higher scores than the traditional DAMP and BCS-SPL, since it enjoys both the benefit of full-image smoothing in BCS-SPL and the idea of high-performance denoising in DAMP. We notice that GSR algorithm even outperforms all network-based approaches on Set11 at the particular sampling rate of 0.4. This is another indication that network-based methods generally offer more advantages over relatively lower sampling rates, while the well-designed traditional approach can achieve excellent performance at very high sampling rates. It can be observed from TABLE I and II that, the competing methods often have higher \textit{PSNR} and \textit{SSIM} values than the original papers. This indicates that those methods can benefit from different sampling rates of the blocks. However, in the existing network-based methods, a block-wise CS sampling procedure has to map a separate network model, and the inter-block correlation can not be utilized, decreasing the quality of the recovered images because of blocking artifacts. That is, they do not take enough advantage of the different sampling rates of blocks.

\begin{table*}[!htb]
\centering
\caption{The detailed quality scores of recovered images in terms of \textit{PRNR} and \textit{SSIM} on the images in Set5. The best performance is labeled in bold.}		
\begin{tabular}{*{9}{c|}c} 		
		\hline
		\multirow{2}*{Images} &Sampling &\multirow{2}*{BCS-SPL} &\multirow{2}*{DAMP} &\multirow{2}*{GSR} &\multirow{2}*{ReconNet} &\multirow{2}*{I-Recon} &\multirow{2}*{Ista} &\multirow{2}*{\textbf{BCS-DAMP}} & \multirow{2}*{\textbf{BCSnet}}\\
        &rate & & & & & & & &\\
		\hline
		\multirow{7}*{Baby} & 0.01 &18.20/0.4838 &4.46/0.0288 &21.30/0.6618 &21.19/0.5582 &24.22/0.6430 &20.87/0.5778 &23.90/0.6591 &\textbf{26.20}/\textbf{0.6887}\\
		& 0.03 &11.73/0.3662 &6.47/0.3125 &26.32/0.8007 &25.78/0.6799 &29.72/0.8004 &25.36/0.7060 &26.49/0.7575 &\textbf{30.62}/\textbf{0.8246}\\
		& 0.05 &26.85/0.8052 &25.86/0.6940 &27.76/0.8539 &27.37/0.7287 &31.32/0.8420 &26.86/0.7342 &27.95/0.8108  &\textbf{32.22}/\textbf{0.8635}\\
		& 0.1 &29.17/0.8750 &28.41/0.7725 &30.64/0.8665 &29.68/0.7998 &33.48/0.8933 &31.83/0.8620 &30.72/0.8913  &\textbf{34.59}/\textbf{0.9113}\\
		& 0.2 &32.49/0.9405 &31.89/0.8637 &33.79/0.9215 &31.87/0.8657 &35.28/0.9164 &34.90/0.9235 &34.29/0.9584 &\textbf{37.62}/\textbf{0.9556}\\
		& 0.3 &34.36/0.9639 &34.34/0.9141 &35.94/\textbf{0.9806} &33.56/0.8977 &36.76/0.9388 &36.92/0.9503 &36.81/0.9782 &\textbf{39.73}/0.9726\\
		& 0.4 &34.89/0.9730 &34.91/0.9732 &37.82/\textbf{0.9883} &32.84/0.8911 &36.65/0.9461 &37.41/0.9611 &37.38/0.9854 &\textbf{40.29}/0.9798\\
		\hline
		\multirow{7}*{Bird} &0.01 &17.90/0.3885 &8.33/0.0225 &19.62/0.5201 &19.38/0.4609 &22.14/0.5726 &19.00/0.4628 &17.32/0.3981 &\textbf{23.30}/\textbf{0.6310} \\
		&0.03 &21.69/0.5453 &11.56/0.2631 &24.00/0.7221 &23.02/0.5981 &26.69/0.7373 &23.20/0.6538 &22.09/0.5611 &\textbf{27.64}/\textbf{0.7840} \\
		&0.05 &24.02/0.6516 &22.58/0.5637 &27.91/0.8380 &25.35/0.7060 &29.59/0.8391 &27.38/0.7812 &24.94/0.7226 &\textbf{30.96}/\textbf{0.8786} \\
		&0.1 &26.91/0.7490 &25.61/0.6856 &34.11/0.9399 &27.95/0.7891 &32.15/0.8968 &31.12/0.8729 &29.11/0.8384 &\textbf{34.69}/\textbf{0.9348} \\
		&0.2 &31.00/0.8594 &32.02/0.8581 &39.03/0.9750 &31.58/0.8860 &36.29/0.9525 &36.66/0.9520 &37.39/0.9615 &\textbf{39.60}/\textbf{0.9765} \\
		&0.3 &33.57/0.9066 &36.64/0.9518 &41.94/0.9855 &33.53/0.9204 &38.35/0.9698 &40.15/0.9752 &42.07/0.9849 &\textbf{42.49}/\textbf{0.9879} \\
		&0.4 &35.50/0.9362 &37.21/0.9753 &44.25/0.9908 &33.22/0.9308 &38.62/0.9788 &43.10/0.9867 &\textbf{44.42}/0.9917 &44.28/\textbf{0.9926} \\
		\hline
		\multirow{7}*{Butterfly} &0.01 &11.98/0.2601 &5.23/0.0006 &11.91/0.2549 &13.36/0.2766 &15.47/0.3802 &12.97/0.2851 &6.04/0.0520 &\textbf{15.71}/\textbf{0.4052}\\
		& 0.03 &14.55/0.3658 &10.05/0.2075 &15.03/0.4451 &16.21/0.4306 &20.16/0.6374 &16.05/0.4499 &14.79/0.4409 &\textbf{20.93}/\textbf{0.6906}\\
		& 0.05 &16.42/0.4391 &14.81/0.4291 &17.37/0.6002 &18.05/0.5404 &22.21/0.7453 &19.94/0.6843 &17.17/0.5946 &\textbf{23.88}/\textbf{0.8160}\\
		& 0.1 &18.84/0.5496 &18.63/0.6547 &22.83/0.8395 &20.58/0.6621 &23.44/0.7666 &25.30/0.8401 &22.58/0.8035 &\textbf{27.93}/\textbf{0.8954}\\
		& 0.2 &21.73/0.6568 &23.78/0.8254 &29.29/\textbf{0.9398} &23.49/0.7688 &26.59/0.8464 &30.40/0.9189 &26.90/0.8902 &\textbf{30.97}/0.9360\\
		& 0.3 &23.94/0.7220 &27.79/0.9081 &32.74/0.9616 &25.41/0.8320 &29.68/0.9183 &\textbf{34.36}/0.9607 &30.08/0.9378 &33.81/\textbf{0.9634}\\
		& 0.4 &25.26/0.7698 &29.77/0.9387 &35.26/0.9731 &25.81/0.8401 &30.34/0.9362 &\textbf{36.50}/0.9741 &33.17/0.9640 &35.50/\textbf{0.9761}\\
		\hline
		\multirow{7}*{Head} &0.01 &17.60/0.3134 &9.44/0.0964 &24.99/0.5970 &21.33/0.5025 &25.30/0.6133 &21.08/0.5199 &24.38/0.4845 &\textbf{28.14}/\textbf{0.6568} \\
		& 0.03 &27.21/0.6312 &24.90/0.4737 &28.04/0.6791 &27.43/0.6080 &30.24/0.7129 &27.69/0.6429 &27.35/0.5828 &\textbf{30.78}/\textbf{0.7369} \\
		& 0.05 &28.36/0.6734 &26.51/0.5368 &29.61/0.7267 &28.83/0.6737 &31.45/0.7583 &29.51/0.7014 &28.56/0.6773 &\textbf{32.03}/\textbf{0.7833} \\
		& 0.1 &30.15/0.7253 &28.69/0.6151 &31.37/0.7853 &30.27/0.7238 &33.05/0.8205 &31.55/0.7713 &30.51/0.7354 &\textbf{33.83}/\textbf{0.8403} \\
		& 0.2 &31.78/0.7870 &31.01/0.7166 &33.70/0.8516 &31.90/0.7818 &34.77/0.8714 &33.49/0.8351 &32.26/0.7850 &\textbf{35.84}/\textbf{0.8965}\\
		& 0.3 &33.37/0.8361 &32.99/0.8086 &35.21/0.8896 &33.04/0.8177 &36.02/0.9008 &35.28/0.8842 &33.72/0.8286 &\textbf{37.53}/\textbf{0.9280}\\
		& 0.4 &34.27/0.8650 &33.92/0.8380 &36.56/0.9167 &32.53/0.8276 &36.02/0.9100 &36.38/0.9117 &34.53/0.8534 &\textbf{38.66}/\textbf{0.9458}\\
		\hline
		\multirow{7}*{Woman} & 0.01 &15.29/0.3608 &5.10/0.0074 &16.60/0.5072 &17.06/0.4475 &20.30/0.5762 &16.41/0.4490 &16.80/0.4497 &\textbf{21.31}/\textbf{0.6370} \\
		& 0.03 &17.94/0.5288 &8.95/0.2337 &21.95/0.7444 &21.03/0.6056 &24.75/0.7494 &21.18/0.6523 &20.07/0.6121 &\textbf{25.36}/\textbf{0.7912} \\
		& 0.05 &22.05/0.6564 &19.66/0.5472 &23.00/0.8195 &22.95/0.6872 &27.12/0.8263 &24.90/0.7648 &23.70/0.7613 &\textbf{28.41}/\textbf{0.8689} \\
		& 0.1 &24.83/0.7506 &24.02/0.7367 &28.87/0.9151 &25.96/0.7842 &29.27/0.8710 &29.75/0.8849 &28.45/0.8715 &\textbf{32.18}/\textbf{0.9234} \\
		& 0.2 &28.32/0.8351 &30.33/0.8852 &33.39/0.9607 &28.89/0.8718 &32.69/0.9250 &34.23/0.9477 &33.74/0.9463 &\textbf{36.16}/\textbf{0.9661} \\
		& 0.3 &30.65/0.8803 &34.50/0.9538 &36.49/0.9762 &30.45/0.9011 &34.53/0.9507 &36.66/0.9689 &37.08/0.9741 &\textbf{38.54}/\textbf{0.9807} \\
		& 0.4 &31.64/0.9053 &35.46/0.9722 &38.75/0.9844 &30.59/0.9068 &34.90/0.9615 &37.32/0.9773 &37.58/0.9836 &\textbf{39.33}/\textbf{0.9864} \\
		\hline
	\end{tabular}
	\label{tab:set5psnr}
\end{table*}

By referring to \cite{johnson2016perceptual}, we define a perceptual feature metric for indicating the high-level perceptual similarity to constrain the recovered semantic information, and the perceptual feature is extracted from a deep network. In the experiments, we employ the 16-layer VGG network pretrained on widely used CIFAR-10 dataset to extract the perceptual features of recovered images at layer relu3\_3. The relative squared Euclidean distance of perceptual features between the ground truth and the recovered image, $\frac{\parallel \phi_j(x)-\phi_j(\hat{x})\parallel_2^2}{\parallel \phi_j(x)\parallel_2^2}$, is used as the \textit{Perceptual Similarity} index, where $\phi_j(\cdot)$ is a feature map of the $j^{th}$ layer. The average quality scores of recovered images in terms of \textit{Perceptual Similarity} on Set5 are illustrated in Table \ref{tab:set5-ps}. It is observed from Table \ref{tab:set5-ps} that, our proposed BCSnet achieves the highest scores according to the mean of the average scores at all seven sampling rates. Table \ref{tab:set5-ps} also shows that the GSR approach obtains excellent performance on the sampling rates of 0.3 and 0.4. This indicates the traditional CS recovery approaches may behave better than network based scheme in the case of relatively higher sampling rates. However, they generally take too much time to recover an image.

\begin{table*}[!htb]
\footnotesize
\centering
\caption{The average quality scores of recovered images in terms of \textit{Perceptual Similarity} on Set5. Lower score indicates better quality. The best performance is labeled in bold.}		\begin{tabular}{c|p{1.3cm}<{\centering}p{1cm}<{\centering}p{1cm}<{\centering}p{1.3cm}<{\centering}p{1.1cm}<{\centering}
p{0.9cm}<{\centering}p{1.5cm}<{\centering}p{0.9cm}<{\centering}}
        \hline
		Sampling rate &BCS-SPL &DAMP &GSR &ReconNet &I-Recon &Ista &BCS-DAMP &BCSnet\\
		\hline
		0.01 &0.9407 &1.1461 &0.8475 &0.9428 &0.8754 &0.9460 &0.8540 &\textbf{0.8293}\\
		0.03 &0.9270 &1.0617 &0.7968 &0.8651 &0.8000 &0.8475 &0.8213 &\textbf{0.7496}\\
		0.05 &0.8432 &0.8912 &0.7542 &0.8382 &0.7579 &0.7974 &0.7924 &\textbf{0.6986}\\
		0.1 &0.7944 &0.8257 &0.6663 &0.7855 &0.7182 &0.6834 &0.7218 &\textbf{0.6191}\\
		0.2 &0.7288 &0.7090 &0.5423 &0.7109 &0.6217 &0.5713 &0.6144 &\textbf{0.5346}\\
		0.3 &0.6736 &0.6117 &\textbf{0.4699} &0.6707 &0.5639 &0.4880 &0.5298  &0.4704\\
		0.4 &0.6473 &0.5475 &\textbf{0.4072} &0.6766 &0.5516 &0.4445 &0.4735  &0.4344\\
		\hline
        Mean &0.7936 &0.8276 &0.6406 &0.7843 &0.6984 &0.6826 &0.6868 &\textbf{0.6194}\\
		\hline
	\end{tabular}
	\label{tab:set5-ps}
\end{table*}

For subjective evaluation, we use MOS to measure the quality of recovered images. The perceived image quality is graded on five levels, from 1 to 5 (bad, poor, fair, good and excellent). In the experiments, 30 volunteers with normal vision, including 10 postgraduates and 20 undergraduates, are involved in the evaluation procedure of the recovered images with eight algorithms. The average quality scores of recovered images in terms of \textit{MOS} on Set5 are illustrated in Table \ref{tab:set5-mos}, which shows that our BCSnet achieves better subjective assessment than the competing methods. We can also see from Table \ref{tab:set5-mos} that, when sampling rates are more than 0.3, GSR and Ista approaches also have relatively high MOS scores. At this time, all the images are well-recovered with these two methods, and the minor difference of the visual quality is hard to identify by the volunteers.

\begin{table*}[!htb]
\footnotesize
\centering
\caption{The average quality scores of recovered images in terms of \textit{MOS} on Set5. Higher score indicates better quality. The best performance is labeled in bold.}		
\begin{tabular}{c|p{1.3cm}<{\centering}p{1cm}<{\centering}p{1cm}<{\centering}p{1.3cm}<{\centering}p{1.1cm}<{\centering}
p{0.9cm}<{\centering}p{1.5cm}<{\centering}p{0.9cm}<{\centering}}
        \hline
		Sampling rate &BCS-SPL &DAMP &GSR &ReconNet &I-Recon &Ista &BCS-DAMP &BCSnet\\
		\hline
		0.01 &1.1313 &1.0071 &1.9709 &1.5599 &2.3269 &1.5396 &1.5797 &\textbf{2.8672}\\
		0.03 &1.7322 &1.2695 &2.6417 &2.3937 &3.5007 &2.5670 &2.5542 &\textbf{4.1310}\\
		0.05 &2.3182 &1.9744 &3.2687 &2.7234 &3.8214 &3.1695 &2.9515 &\textbf{4.5704}\\
		0.1 &2.9657 &2.7669 &4.2933 &3.2434 &4.1256 &4.0413 &3.6735 &\textbf{4.7665}\\
		0.2 &3.6405 &3.8165 &4.5161 &3.8318 &4.3613 &4.5548 &4.4170 &\textbf{4.8258}\\
		0.3 &4.0779 &4.4065 &4.6626 &4.2291 &4.5652 &4.6822 &4.4899 &\textbf{4.8387}\\
		0.4 &4.2026 &4.5378 &4.6875 &4.2744 &4.6839 &4.6955 &4.6323 &\textbf{4.8452}\\
		\hline
	\end{tabular}
	\label{tab:set5-mos}
\end{table*}

The average running time reconstructing the images in Set5 with the target sampling rate of 0.1 is shown in Table~\ref{tab:time}.
Here, the time of BCS-SPL, GSR, and BCS-DAMP is the average of reconstructing four images in Set5 without ``Baby" of large size $512\times 512$, considering it takes too long to complete the reconstruction, and the time of DAMP is that of reconstructing only one block, since all blocks may be easily processed in parallel. It is observed from Table~\ref{tab:time} that, BCS-SPL and our BCS-DAMP have much longer running time than DAMP algorithm due to the full-image processing strategy. As mentioned above, in the experiment GSR is with fixed-rate allocation, not adaptive sampling like BCS-SPL and DAMP. However, even with the simple fixed sampling rate, GSR still needs 1089 seconds to recover four images in Set5 on average in order to obtain a good quality. That is, although the different sampling rates of blocks are known to GSR algorithm, GSR can not utilize them because of overlong running time. The proposed BCSnet runs a little longer than network-based ReconNet, I-Recon and Ista because of the multi-channel sampling and block-wise reassembling. But it runs significantly faster than traditional BCS-SPL, DAMP, GSR, and BCS-DAMP reconstruction algorithms. We should notice that, the assignment of different sampling rate requires an initial sensing, which introduces an additional time consumption. Here we omit it with no impact on the comparison, since our scheme and the competing methods employ the same adaptive sampling strategy.

\begin{table*}[!htb]
	\centering
	\caption{The average time for reconstructing the images in Set5 with the sampling rate of 0.1 (in second), where BCS-SPL, GSR, and BCS-DAMP are the average of four images in Set5 without ``Baby".}
	\begin{tabular}{c|ccccccccc}
		\hline
		Algorithm &BCS-SPL &DAMP &GSR &BCS-DAMP &ReconNet &I-Recon &Ista &BCSnet-WBA &BCSnet\\
		\hline
		Time &471s &20.63s &1089s &323s &0.88s &0.90s &0.99s &1.59s &2.99s \\
		\hline
	\end{tabular}
	\label{tab:time}
\end{table*}

\subsubsection{Performance evaluation with fixed-rate allocation}

The performance of our scheme is further evaluated in the case with fixed sampling rate (FSR). That is, all blocks within an image are with the same sampling rate, since the sensing resources may not be allocated in certain scenarios. Accordingly, they are fed into the model via the same channel, and the proposed multi-channel structure then becomes a unified deep network with $k$ target sampling rates. We term this version of the BCSnet as BCSnet-FSR. In this simulation, $k=7$ is taking as an example, and the channels fall in set $\{0.01, 0.03, 0.05, 0.1, 0.2, 0.3, 0.4\}$.

\begin{table*}[!htb]
	\centering
	\caption{The average \textit{PSNR} and \textit{SSIM} with FSR strategy. The best performance scores are labeled in bold.}
	\begin{tabular}{c|ccccccc}
		\hline
		\multicolumn{8}{c}{Set5 (\textit{PSNR}/\textit{SSIM})}\\
		\hline
		\diagbox[width=10em]{Algorithm}{Sampling rate} &0.01 &0.03 &0.05 &0.1 &0.2 &0.3 &0.4\\
		\hline
		\multirow{2}*{Existing algorithm} &21.49/0.5571 &25.00/0.7113 &26.97/0.7908 &29.56/0.8692 &33.84/0.9297 &36.46/0.9587 &38.53/0.9707\\
        &(I-Recon-FSR) &(I-Recon-FSR) &(I-Recon-FSR) &(GSR) &(GSR) &(GSR) &(GSR)\\
        \hline
		BCS-DAMP-FSR &17.69/0.4087 &22.56/0.6419 &24.12/0.7063 &27.42/0.8191 &32.27/0.9110 &34.99/0.9385 &37.42/0.9556\\
		BCSnet-WBA-FSR &22.52/0.5844 &26.04/0.7378 &27.92/0.8097 &30.77/0.8834 &34.41/0.9391 &36.92/0.9616 &38.54/0.9723\\
		BCSnet-FSR &\textbf{22.93}/\textbf{0.6037} &\textbf{26.63}/\textbf{0.7655} &\textbf{28.69}/\textbf{0.8342} &\textbf{31.81}/\textbf{0.9012} &\textbf{35.32}/\textbf{0.9463} &\textbf{37.69}/\textbf{0.9654} &\textbf{39.61}/\textbf{0.9762}\\
		\hline
		\multicolumn{8}{c}{Set11 (\textit{PSNR}/\textit{SSIM})}\\
		\hline
		\diagbox[width=10em]{Algorithm}{Sampling rate} &0.01 &0.03 &0.05 &0.1 &0.2 &0.3 &0.4\\
		\hline
		\multirow{2}*{Existing algorithm} &19.80/0.5018 &22.59/0.6540 &24.54/0.7442 &26.49/0.8010 &\textbf{32.15}/\textbf{0.9338} &\textbf{34.75}/0.9465 &\textbf{36.89}/\textbf{0.9693}\\
        &(I-Recon-FSR) &(I-Recon-FSR) &(I-Recon-FSR) &(Ista-FSR) &(GSR) &(GSR) &(GSR)\\
        \hline
		BCS-DAMP-FSR &18.13/0.4781 &20.81/0.5811 &22.38/0.6503 &25.76/0.7887 &30.19/0.8977 &33.22/0.9401 &35.60/0.9596\\
		BCSnet-WBA-FSR &20.52/0.5241 &23.42/0.6727 &25.19/0.7567 &27.71/0.8409 &31.01/0.9087 &33.61/0.9419 &35.44/0.9583\\
		BCSnet-FSR &\textbf{20.81}/\textbf{0.5427} &\textbf{23.97}/\textbf{0.7003} &\textbf{25.85}/\textbf{0.7819} &\textbf{28.52}/\textbf{0.8599} &31.92/0.9190 &34.46/\textbf{0.9480} &36.52/0.9640\\
		\hline
		\multicolumn{8}{c}{BSD100 (\textit{PSNR}/\textit{SSIM})}\\
		\hline
		\diagbox[width=10em]{Algorithm}{Sampling rate} &0.01 &0.03 &0.05 &0.1 &0.2 &0.3 &0.4\\
		\hline
		\multirow{2}*{Existing algorithm} &21.15/0.4654 &23.09/0.5662 &24.19/0.6320 &25.35/0.7098 &27.64/0.7906 &29.86/0.8593 &31.70/0.9003\\
        &(I-Recon-FSR) &(I-Recon-FSR) &(I-Recon-FSR) &(I-Recon-FSR) &(Ista-FSR) &(Ista/I-Recon-FSR) &(Ista-FSR)\\
        \hline
		BCS-DAMP-FSR &19.41/0.4199 &21.33/0.4899 &22.34/0.5253 &23.82/0.5849 &26.24/0.6815 &28.12/0.7533 &30.21/0.8241\\
		BCSnet-WBA-FSR &21.75/0.4836 &23.77/0.5900 &24.82/0.6527 &26.52/0.7479 &29.00/0.8481 &31.03/0.9015 &32.71/0.9319\\
		BCSnet-FSR &\textbf{21.97}/\textbf{0.4943} &\textbf{24.03}/\textbf{0.6049} &\textbf{25.11}/\textbf{0.6684} &\textbf{26.93}/\textbf{0.7632} &\textbf{29.44}/\textbf{0.8588} &\textbf{31.49}/\textbf{0.9084} &\textbf{33.33}/\textbf{0.9383}\\
		\hline
	\end{tabular}
	\label{tab:compa-was}
\end{table*}

The comparison of average \textit{PSNR} and \textit{SSIM} of recovered Set5, Set11, and BSD100 is indicated in Table \ref{tab:compa-was}, where ``existing algorithm" denotes the one, including BCS-SPL, DAMP, GSR, ReconNet, I-Recon and Ista, which achieves the maximal value of \textit{PSNR} or \textit{SSIM}. One can see that, in the application scenario that each block has to be assigned the same sampling rate, BCSnet-FSR still achieves higher \textit{PSNR} and \textit{SSIM} than the existing algorithms. Observing Table \ref{tab:compa-was}, except for the PSNR scores of Set11 with high sampling rates of 0.2, 0.3, and 0.4, the weaken version of BCSnet-FSR, BCSnet-WBA-FSR, also has the better performance than the existing algorithm for every dataset, due to the full-image denoising strategy. The performance of BCSnet-WBA-FSR is further improved by BCSnet-FSR, where the technique of block-wise approximation is added to improve the recovered images. Compared with BCSnet whose quality scores are shown in Table~\ref{tab:compa}, BCSnet-FSR always also has relatively lower average \textit{PSNR} with the sampling rates of $\{0.03, 0.05, 0.1, 0.2, 0.3\}$ for the images in each set, since the fixed sampling rates is often less efficient than adaptive allocation. One can also see that BCSnet-FSR has the same \textit{PSNR} and \textit{SSIM} as BCSnet with the target rates of 0.01 and 0.4, since 0.01 and 0.4 are the boundary sampling rates, and the adaptive allocation strategy degenerates to uniform allocation way adopted in BCSnet-FSR.

In the application scenarios with fixed sampling rates, the proposed $k$-channel sampling structure can be used to serve all $k$ target sampling rates with a unified deep reconstruction network. A large amount of storage space is then be saved. For instance, Ista approach for seven different target sampling rates has about 0.34 million (M) parameters each, requiring 2.38M in all. In contrast, our reconstruction network has only 0.75M parameters due to the fact that $k$ target sampling rates are integrated into a single model.

\begin{figure*}[!htb]
\centering
\includegraphics[width=2.1\OneImW]{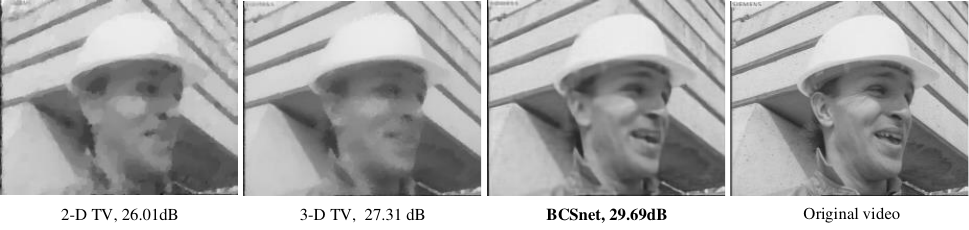}
\caption{A video recovery comparison with different models at the sampling rate of 0.05. Here we provide the $10^{th}$ frame of the recovered video and the average \textit{PSNR} scores of all 16 frames in the video.}
\label{fig:visual005}
\end{figure*}

\begin{figure*}[!htb]
\centering
\includegraphics[width=2.1\OneImW]{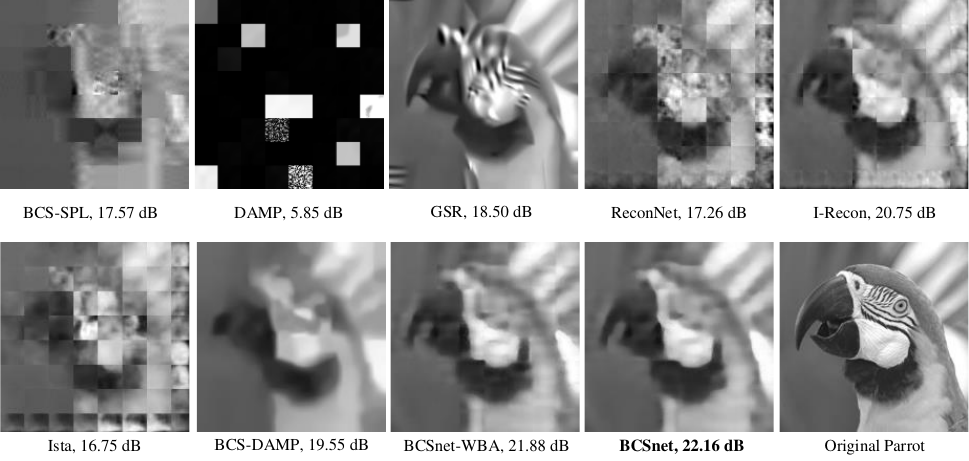}
\caption{Visual effects of the recovered ``Parrot" with extremely low sampling rate of 0.01.}
\label{fig:visual001}
\end{figure*}

\begin{figure*}[!htb]
\centering
\includegraphics[width=2.1\OneImW]{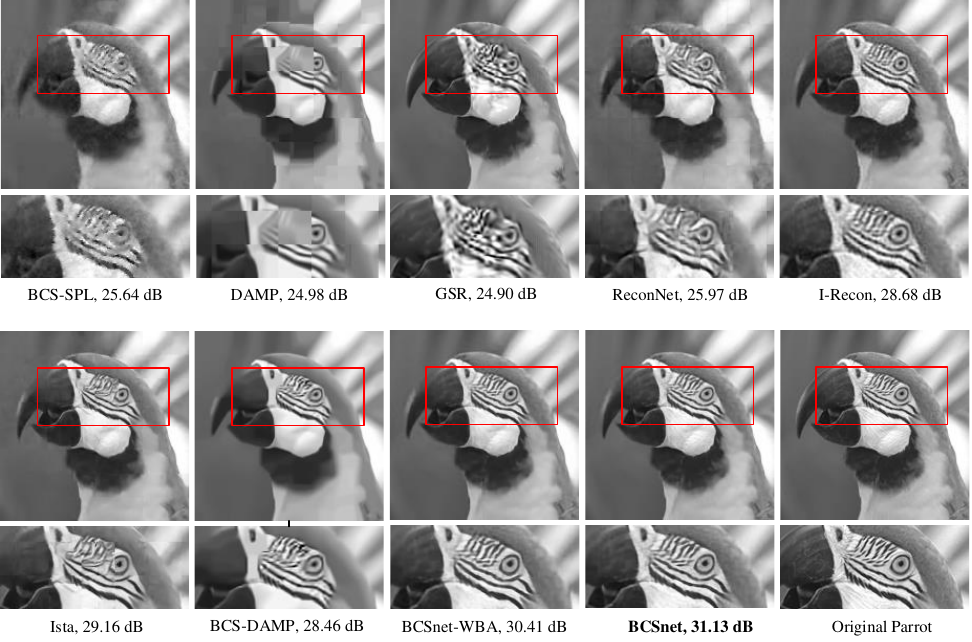}
\caption{Visual effects of the recovered ``Parrot" with sampling rate of 0.1.}
\label{fig:visual01}
\end{figure*}

\subsubsection{Comparison of visual effects}

Using 16-frame video ``Foreman" and image ``Parrot" as test datasets, this section compares the visual effect of the recovered videos and images with our methods and other competing ones.

The video recovery comparison with different video models at the sampling rate of 0.05 is illustrated in Fig.~\ref{fig:visual005}. In the experiment, the saliency map of the first frame is first computed, and this map is used by the following 15 frames to assign the sampling rates. The assignment of the different sampling rates are also used by the competing video models for a fair comparison. For our BCSnet and the 2-D TV model, the video is recovered frame by frame. For 3-D TV model, the video recovery is achieved by stacking 16 frames into a whole and employ 3-D discrete gradient across spatial dimensions and time. Fig.~\ref{fig:visual005} shows that our scheme outperforms 2-D TV and 3-D TV video models by a large margin in terms of visual quality and \textit{PSNR} scores, thanks to the powerful learning capacity of the proposed deep network architecture.

It can be observed from Fig.~\ref{fig:visual001} that, in the case of extremely low sampling rate of 0.01, all methods achieve the recovered images with low quality. However, the proposed BCSnet has relatively better visual effect than the competing ones. It is observed that the traditional BCS-SPL and DAMP, due to too few measurements with the sampling rate of 0.01, almost fail to reconstruct the images. The proposed BCS-DAMP borrows the ideas of block-wise approximation and full-image denoising from these two methods, and outperforms them in terms of visual effect and quality scores. For network-based ReconNet, I-Recon, and Ista methods, one can easily observe significant blocking artifacts because of their block-by-block reconstruction strategy. In contrast, our methods, including the weak BCSnet-WBA version, are capable of reconstructing smoother images, and nearly removing the blocking artifact incurred by block-wise CS sampling.

From Fig.~\ref{fig:visual01}, one can see that the recovered ``Parrot" are all improved when the sampling rate increases to 0.1. With the traditional DAMP and networked ReconNet, the blocking artifacts can still be observed, while with GSR, I-Recon, and Ista, the blocking artifacts are relatively less obvious. However, carefully observing the parts of the eye and the beak in the recovered ``Parrot", the proposed scheme is capable of reconstructing more details, sharper edges, and significantly less blocking artifacts than the competing ones. That is, our BCSnet perfectly synthesizes the common merits of BCS-SPL algorithm and deep network approach, and achieves the best performance.

\section{Conclusion}

In this paper, we further studied the problem of block-based image compressive sensing, and proposed a multi-channel deep neural network structure, termed `BCSnet'. Based on the popular BCS-SPL algorithm, block-wise iterative approximation and denoising in the scope of full image were iteratively performed to improve the recovered image. Furthermore, the framework was cast into a carefully designed deep network, so that BCSnet is capable of benefitting both from the learning capacities of deep network and the hand-designed structure of BCS-SPL algorithm. Extensive experimental results show that BCSnet with adaptive sensing resource allocation achieves far better reconstruction quality and superb visual effect compared with state-of-the-art methods. In addition, BCSnet without the adaptive allocation strategy also has excellent reconstruction performance with significantly less network parameters, which extends the application scenarios of the proposed method of image recovery.

\bibliographystyle{IEEEtran_doi}
\bibliography{deepcs}

\begin{thebibliography}{10}
\providecommand{\url}[1]{#1}
\csname url@samestyle\endcsname
\providecommand{\newblock}{\relax}
\providecommand{\bibinfo}[2]{#2}
\providecommand{\BIBentrySTDinterwordspacing}{\spaceskip=0pt\relax}
\providecommand{\BIBentryALTinterwordstretchfactor}{4}
\providecommand{\BIBentryALTinterwordspacing}{\spaceskip=\fontdimen2\font plus
\BIBentryALTinterwordstretchfactor\fontdimen3\font minus
  \fontdimen4\font\relax}
\providecommand{\BIBforeignlanguage}[2]{{%
\expandafter\ifx\csname l@#1\endcsname\relax
\typeout{** WARNING: IEEEtran.bst: No hyphenation pattern has been}%
\typeout{** loaded for the language `#1'. Using the pattern for}%
\typeout{** the default language instead.}%
\else
\language=\csname l@#1\endcsname
\fi
#2}}
\providecommand{\BIBdecl}{\relax}
\BIBdecl

\bibitem{CS}
D.~Donoho, ``\href{http://dx.doi.org/10.1109/TIT.2006.871582}{Compressed
  sensing},'' \emph{IEEE Transactions on Information Theory}, vol.~52, no.~4,
  pp. 1289--1306, 2006.

\bibitem{elad2007optimized}
M.~Elad, ``Optimized projections for compressed sensing,'' \emph{IEEE
  Transactions on Signal Processing}, vol.~55, no.~12, pp. 5695--5702, 2007.

\bibitem{CS-MM1}
Q.~Jiang, S.~Li, Z.~Zhu, H.~Bai, X.~He, and R.~C. de~Lamare, ``Design of
  compressed sensing system with probability-based prior information,''
  \emph{IEEE Transactions on Multimedia (Early Access)}, 2019.

\bibitem{singlecamera}
M.~F. Duarte, M.~A. Davenport, D.~Takhar, J.~N. Laska, T.~Sun, K.~F. Kelly, and
  R.~G. Baraniuk,
  ``\href{http://dx.doi.org/10.1109/MSP.2007.914730}{Single-pixel imaging via
  compressive sampling},'' \emph{IEEE Signal Processing Magazine}, vol.~25,
  no.~2, pp. 83--91, 2008.

\bibitem{MRI}
M.~Lustig, D.~L. Donoho, J.~M. Santos, and J.~M. Pauly,
  ``\href{http://dx.doi.org/10.1109/MSP.2007.914728}{Compressed sensing
  {MRI}},'' \emph{IEEE Signal Processing Magazine}, vol.~25, no.~2, pp. 72--82,
  2008.

\bibitem{Trans-MM-1}
Z.~Chen, X.~Hou, X.~Qian, and C.~Gong, ``Efficient and robust image coding and
  transmission based on scrambled block compressive sensing,'' \emph{IEEE
  Transactions on Multimedia}, vol.~20, no.~7, pp. 1610--1621, 2017.

\bibitem{Trans-MM-2}
S.~Zheng, X.-P. Zhang, J.~Chen, and Y.~Kuo, ``A high-efficiency compressed
  sensing based terminal-to-cloud video transmission system,'' \emph{IEEE
  Transactions on Multimedia}, vol.~21, no.~8, pp. 1905--1920, 2019.

\bibitem{cqli:meet:JISA19}
C.~Li, Y.~Zhang, and E.~Y. Xie,
  ``\href{http://dx.doi.org/10.1016/j.jisa.2019.102361}{{When an attacker meets
  a cipher-image in 2018: A Year in Review}},'' \emph{Journal of Information
  Security and Applications}, vol.~48, p. art. no. 102361, 2019.

\bibitem{BCS}
L.~Gan, ``Block compressed sensing of natural images,'' in \emph{Proceedings of
  the International Conference on Digital Signal Processing}, 2007, pp.
  403--406.

\bibitem{SPL}
J.~Bigot, C.~Boyer, and P.~Weiss, ``An analysis of block sampling strategies in
  compressed sensing,'' \emph{IEEE transactions on information theory},
  vol.~62, no.~4, pp. 2125--2139, 2016.

\bibitem{BCS-1}
J.~Fowler, {Sungkwang Mun}, and E.~Tramel,
  ``\href{http://dx.doi.org/10.1561/2000000033}{Block-based compressed sensing
  of images and video},'' \emph{Foundations and Trends in Signal Processing},
  vol.~4, no.~4, pp. 297--416, 2010.

\bibitem{BCS-Salie}
Y.~Yu, B.~Wang, and L.~Zhang, ``Saliency-based compressive sampling for image
  signals,'' \emph{IEEE Signal Processing Letters}, vol.~17, no.~11, pp.
  973--976, 2010.

\bibitem{artifact}
C.~Zhao, J.~Zhang, S.~Ma, X.~Fan, Y.~Zhang, and W.~Gao, ``Reducing image
  compression artifacts by structural sparse representation and quantization
  constraint prior,'' \emph{IEEE Transactions on Circuits and Systems for Video
  Technology}, vol.~27, no.~10, pp. 2057--2071, 2017.

\bibitem{DNN-C}
A.~Krizhevsky, I.~Sutskever, and G.~E. Hinton,
  ``\href{http://dx.doi.org/10.1145/3065386}{Imagenet classification with deep
  convolutional neural networks},'' \emph{Commnications of the ACM}, vol.~60,
  no.~6, pp. 84--90, 2017.

\bibitem{galteri2019deep}
L.~Galteri, L.~Seidenari, M.~Bertini, and A.~Del~Bimbo, ``Deep universal
  generative adversarial compression artifact removal,'' \emph{IEEE
  Transactions on Multimedia}, vol.~21, no.~8, pp. 2131--2145, 2019.

\bibitem{Reconnet}
K.~Kulkarni, S.~Lohit, P.~Turaga, R.~Kerviche, and A.~Ashok, ``Reconnet:
  Non-iterative reconstruction of images from compressively sensed
  measurements,'' in \emph{IEEE Conference on Computer Vision and Pattern
  Recognition (CVPR)}, 2016, pp. 449--458.

\bibitem{Ldamp}
C.~Metzler, A.~Mousavi, and R.~Baraniuk, ``Learned {D-AMP}: Principled neural
  network based compressive image recovery,'' in \emph{Proceedings of the 31st
  International Conference on Neural Information Processing Systems (NIPS)},
  2017, pp. 1770--1781.

\bibitem{Im-recon}
S.~Lohit, K.~Kulkarni, R.~Kerviche, P.~Turaga, and A.~Ashok, ``Convolutional
  neural networks for non-iterative reconstruction of compressively sensed
  images,'' \emph{IEEE Transactions on Computational Imaging}, vol.~4, no.~3,
  pp. 326--340, 2018.

\bibitem{ISTA}
J.~Zhang and B.~Ghanem, ``Ista-net: Interpretable optimization-inspired deep
  network for image compressive sensing,'' in \emph{IEEE Conference on Computer
  Vision and Pattern Recognition (CVPR)}, 2018, pp. 1828--1837.

\bibitem{DnCNN}
K.~Zhang, W.~Zuo, Y.~Chen, D.~Meng, and L.~Zhang,
  ``\href{http://dx.doi.org/10.1109/TIP.2017.2662206}{Beyond a gaussian
  denoiser: Residual learning of deep cnn for image denoising},'' \emph{IEEE
  Transactions on Image Processing}, vol.~26, no.~7, pp. 3142--3155, 2017.

\bibitem{BP}
S.~S. Chen, D.~L. Donoho, and M.~A. Saunders,
  ``\href{http://dx.doi.org/10.1137/S1064827596304010}{Atomic decomposition by
  basis pursuit},'' \emph{SIAM review}, vol.~43, no.~1, pp. 129--159, 2001.

\bibitem{OMP}
M.~Yang and F.~de~Hoog,
  ``\href{http://dx.doi.org/10.1109/TSP.2015.2453137}{Orthogonal matching
  pursuit with thresholding and its application in compressive sensing},''
  \emph{IEEE Transactions on Signal Processing}, vol.~63, no.~20, pp.
  5479--5486, 2015.

\bibitem{GSR}
J.~Zhang, D.~Zhao, and W.~Gao, ``Group-based sparse representation for image
  restoration,'' \emph{IEEE Transactions on Image Processing}, vol.~23, no.~8,
  pp. 3336--3351, 2014.

\bibitem{dong2014compressive}
W.~Dong, G.~Shi, X.~Li, Y.~Ma, and F.~Huang, ``Compressive sensing via nonlocal
  low-rank regularization.'' \emph{IEEE Transactions on Image Processing},
  vol.~23, no.~8, pp. 3618--3632, 2014.

\bibitem{Damp}
C.~A. Metzler, A.~Maleki, and R.~G. Baraniuk,
  ``\href{http://dx.doi.org/10.1109/TIT.2016.2556683}{From denoising to
  compressed sensing},'' \emph{IEEE Transactions on Information Theory},
  vol.~62, no.~9, pp. 5117--5144, 2016.

\bibitem{zhu2018on}
Z.~Zhu, G.~Li, J.~Ding, Q.~Li, and X.~He, ``On collaborative compressive
  sensing systems: The framework, design, and algorithm,'' \emph{Siam Journal
  on Imaging Sciences}, vol.~11, no.~2, pp. 1717--1758, 2018.

\bibitem{MTAP}
S.~Zheng, J.~Chen, and Y.~Kuo, ``A multi-level residual reconstruction based
  image compressed sensing recovery scheme,'' \emph{Multimedia Tools and
  Applications}, vol.~78, pp. 25\,101--25\,119, 2019.

\bibitem{chen2017compressive}
J.~Chen, Y.~Gao, C.~Ma, and Y.~Kuo, ``Compressive sensing image reconstruction
  based on multiple regulation constraints,'' \emph{Circuits Systems and Signal
  Processing}, vol.~36, no.~4, pp. 1621--1638, 2017.

\bibitem{BCS-Small}
S.~Mun and J.~E. Fowler, ``Block compressed sensing of images using directional
  transforms,'' in \emph{International Conference on Image Processing (ICIP)},
  2009, pp. 2985--2988.

\bibitem{adaptive}
A.~Akbari, D.~Mandache, M.~Trocan, and B.~Granado, ``Adaptive saliency-based
  compressive sensing image reconstruction,'' in \emph{IEEE International
  Conference on Multimedia \& Expo Workshops (ICMEW)}, 2016, pp. 1--6.

\bibitem{video}
R.~G. Baraniuk, T.~Goldstein, A.~C. Sankaranarayanan, C.~Studer,
  A.~Veeraraghavan, and M.~B. Wakin, ``Compressive video sensing: Algorithms,
  architectures, and applications,'' \emph{IEEE Signal Processing Magazine},
  vol.~34, no.~1, pp. 52--66, 2017.

\bibitem{Asymmetric}
S.~Zhou, S.~Xiang, X.~Liu, and H.~Li,
  ``\href{http://dx.doi.org/10.1109/ICME.2018.8486517}{Asymmetric block based
  compressive sensing for image signals},'' in \emph{IEEE International
  Conference on Multimedia and Expo (ICME)}, 2018, pp. 1--6.

\bibitem{Fullnet}
X.~Xie, C.~Wang, J.~Du, and G.~Shi,
  ``\href{http://dx.doi.org/10.1109/ICME.2018.8486521}{Full image recover for
  block-based compressive sensing},'' in \emph{IEEE International Conference on
  Multimedia and Expo (ICME)}.\hskip 1em plus 0.5em minus 0.4em\relax IEEE,
  2018, pp. 1--6.

\bibitem{Csnet}
W.~Shi, F.~Jiang, S.~Zhang, and D.~Zhao,
  ``\href{http://dx.doi.org/10.1109/ICME.2017.8019428}{Deep networks for
  compressed image sensing},'' in \emph{IEEE International Conference on
  Multimedia and Expo (ICME)}, 2017, pp. 877--882.

\bibitem{Enhanced}
B.~Lim, S.~Son, H.~Kim, S.~Nah, and K.~M. Lee, ``Enhanced deep residual
  networks for single image super-resolution,'' in \emph{IEEE Conference on
  Computer Vision and Pattern Recognition Workshops (CVPRW)}, 2017, pp.
  1132--1140.

\bibitem{Multi}
W.~Shi, F.~Jiang, S.~Liu, and D.~Zhao, ``Multi-scale deep networks for image
  compressed sensing,'' in \emph{IEEE International Conference on Image
  Processing (ICIP)}, 2018, pp. 46--50.

\bibitem{cheng2018model}
Y.~Cheng, D.~Wang, P.~Zhou, and T.~Zhang, ``Model compression and acceleration
  for deep neural networks: The principles, progress, and challenges,''
  \emph{IEEE Signal Processing Magazine}, vol.~35, no.~1, pp. 126--136, 2018.

\bibitem{BM3D}
K.~Dabov, A.~Foi, V.~Katkovnik, and K.~Egiazarian,
  ``\href{http://dx.doi.org/10.1109/TIP.2007.901238}{Image denoising by sparse
  3-{D} transform-domain collaborative filtering},'' \emph{IEEE Transactions on
  Image Processing}, vol.~16, no.~8, pp. 2080--2095, 2007.

\bibitem{Tensor}
M.~Abadi, P.~Barham, J.~Chen, Z.~Chen, A.~Davis, J.~Dean, M.~Devin,
  S.~Ghemawat, G.~Irving, M.~Isard \emph{et~al.}, ``Tensorflow: A system for
  large-scale machine learning,'' in \emph{12th USENIX Symposium on Operating
  Systems Design and Implementation (OSDI)}, 2016, pp. 265--283.

\bibitem{BSD500}
P.~Arbelaez, M.~Maire, C.~Fowlkes, and J.~Malik, ``Contour detection and
  hierarchical image segmentation,'' \emph{IEEE Transactions on Pattern
  Analysis \& Machine Intelligence}, vol.~33, no.~5, pp. 898--916, 2011.

\bibitem{johnson2016perceptual}
J.~Justin, A.~Alexandre, and F.~Li, ``Perceptual losses for real-time style
  transfer and super-resolution,'' in \emph{European conference on computer
  vision (ECCV)}, 2016, pp. 694--711.

\end{thebibliography}

\graphicspath{{author-figures-pdf/}}


\begin{IEEEbiography}[{\includegraphics[width=1.1in,height=1.25in,clip,keepaspectratio]{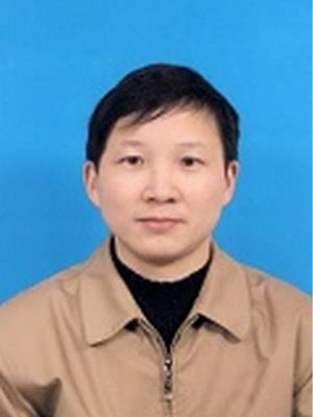}}]{Siwang Zhou}
received his B.S. degree from Fudan University, M.S. degree from Xiangtan University, and the Ph.D. degree from Hunan University, China, respectively. He has been a professor with the College of Computer Science and Electronic Engineering, Hunan University, China. His research interests include image compressive sensing, deep learning, and Internet of Things.
\end{IEEEbiography}

\vskip 0pt plus -1fil

\begin{IEEEbiography}[{\includegraphics[width=1in,height=1.25in,clip,keepaspectratio]{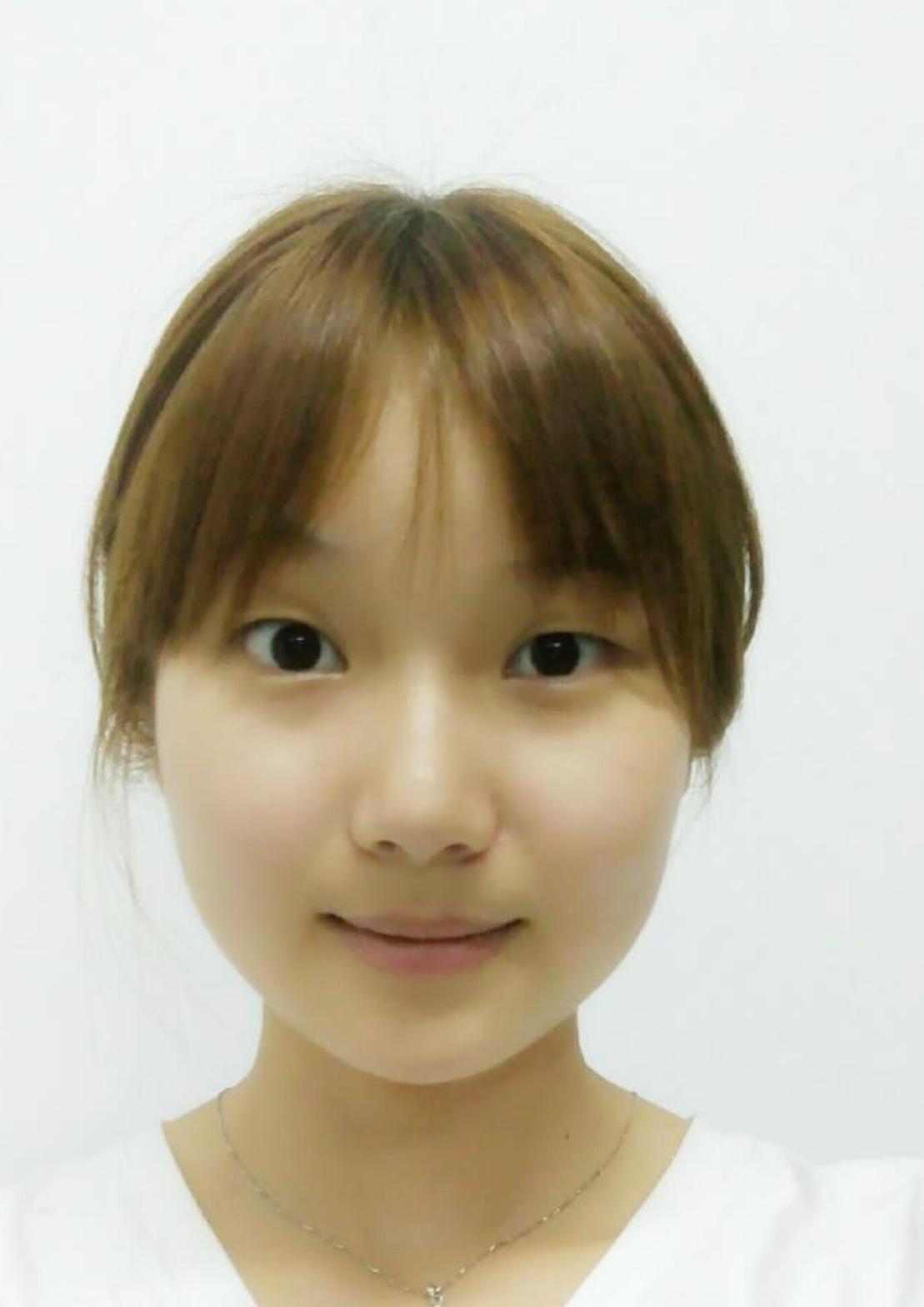}}]{Yan He} is a master student major in software engineering at the College of Computer Science and Electronic Engineering in Hunan University, China. Her research focuses on image compressive sensing.
\end{IEEEbiography}

\vskip 0pt plus -1fil

\begin{IEEEbiography}[{\includegraphics[width=1in,height=1.25in,clip,keepaspectratio]{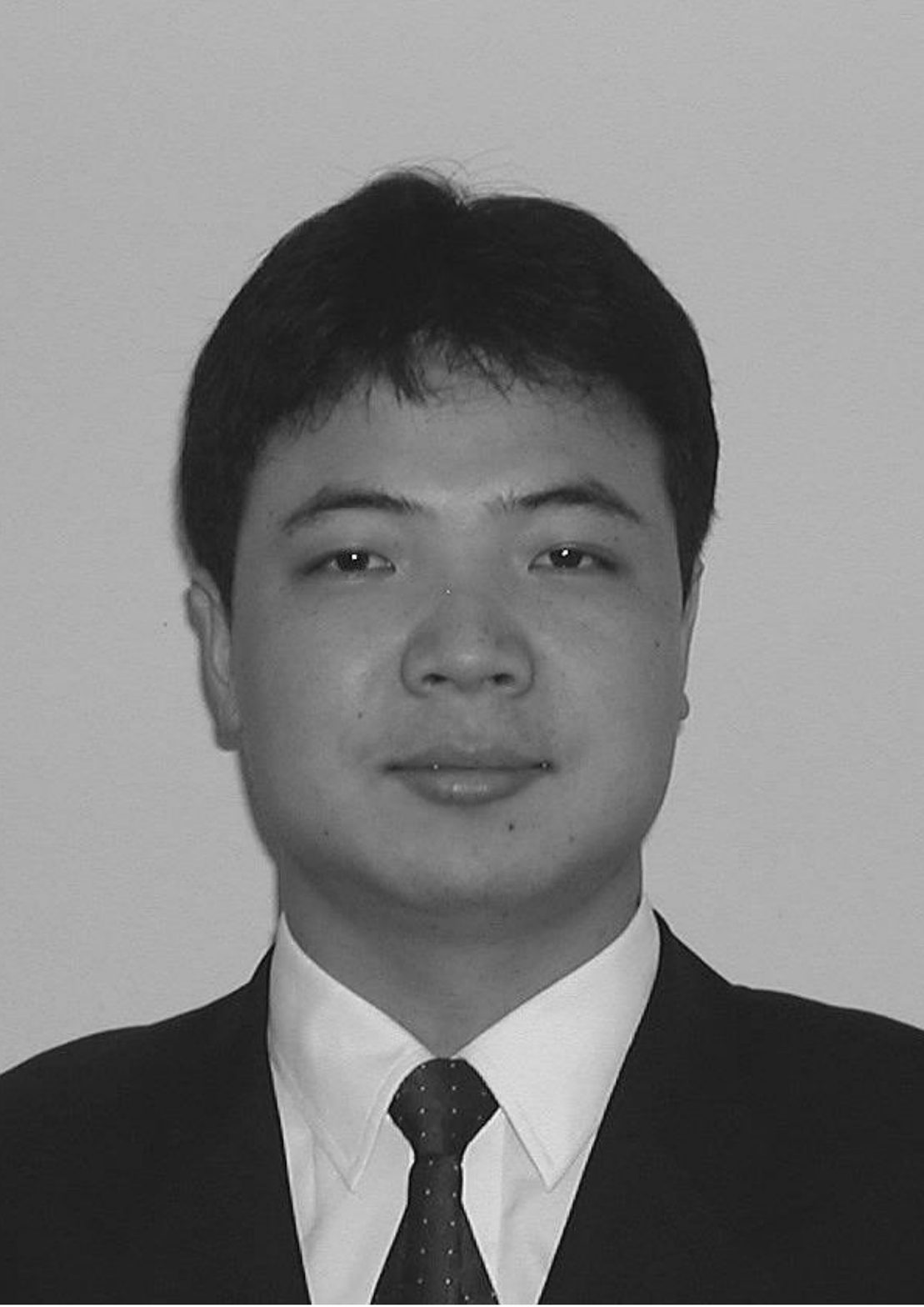}}]{Yonghe Liu} received the B.S. and M.S. degrees from Tsinghua University, Beijing, China, in 1998 and 1999, respectively, and the Ph.D. degree from Rice University, Houston, TX, USA, in 2004. He is an Associate Professor with the Department of Computer Science and Engineering, University of Texas at Arlington, Arlington, TX, USA. His current research interests include wireless networking, sensor networks, security, and system integration.
\end{IEEEbiography}


\vskip 0pt plus -1fil

\begin{IEEEbiography}[{\includegraphics[width=1.1in, height=1.25in,clip,keepaspectratio]{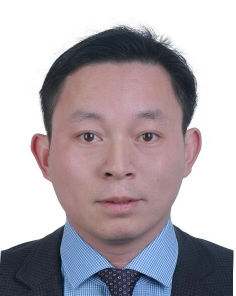}}]{Chengqing Li}(M'07-SM'13)
received his M.Sc. degree in Applied Mathematics from Zhejiang University, China in 2005 and
his Ph.D. degree in Electronic Engineering from City University of Hong Kong in 2008. Thereafter,
he worked as a Postdoctoral Fellow at The Hong Kong Polytechnic University till September 2010.
Then, he worked at the College of Information Engineering, Xiangtan University, China. From April 2013 to July 2014, he worked at the
University of Konstanz, Germany, under the support of the Alexander von Humboldt Foundation.
From May 2020, he have been working at School of Computer Science, Xiangtan University, China as the dean.
He is serving as an associate editor for the International Journal of Bifurcation and Chaos and Signal Processing.

Prof. Li focuses on security analysis of multimedia encryption schemes and privacy protection schemes.
He has published more than fifty papers on the subject in the past 16 years, receiving more than 3700
citations with h-index 34.
\end{IEEEbiography}

\vskip 0pt plus -1fil

\begin{IEEEbiography}[{\includegraphics[width=1in,height=1.25in,clip,keepaspectratio]{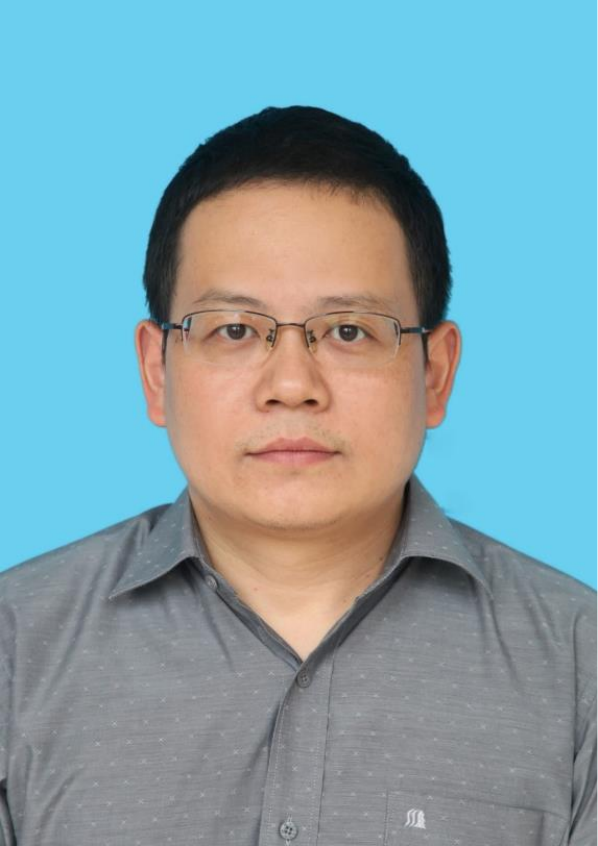}}]{Jianming Zhang} received the B.S. degree from Zhejiang University, in 1996, the M.S. degree from the National University of Defense Technology, China, in 2001, and the Ph.D. degree from Hunan University, China, in 2010. He is currently a Full Professor with the School of Computer and Communication Engineering, Changsha University of Science and Technology, China. His research interests include computer vision, pattern recognition, Internet of Things and mobile computing.
\end{IEEEbiography}


\end{document}